\documentclass[12pt]{article}
\usepackage{amsfonts}
\usepackage{graphicx}
\usepackage{amsmath}
\usepackage{color,soul}
\usepackage{float}
\textwidth=6.5in \hoffset=-.75in \voffset=-1in \numberwithin{equation}{section}
\numberwithin{figure}{section}

\newcommand {\nn}{\nonumber}
\newcommand {\be}{\begin{equation}}
\newcommand {\ee}{\end{equation}}
\newcommand*{\defeq}{\stackrel{\text{def}}{=}}

\begin{document}

\begin{titlepage}
\vspace{1cm}
\begin{center}
{\Large \bf {Bianchi IX geometry and the Einstein-Maxwell theory }}\\
\end{center}
\vspace{2cm}
\begin{center}
A. M. Ghezelbash{ \footnote{ E-Mail: masoud.ghezelbash@usask.ca}}
\\
Department of Physics and Engineering Physics, \\ University of Saskatchewan, \\
Saskatoon, Saskatchewan S7N 5E2, Canada\\
\vspace{1cm}
\vspace{2cm}
\end{center}

\begin{abstract}

We construct numerical solutions to the higher-dimensional Einstein-Maxwell theory. The solutions are based on embedding the four dimensional Bianchi type IX space in the theory. We find the solutions as superposition of  two functions, which one of them can be found numerically. We show that the solutions  in any dimensions, are almost regular everywhere, except a singular point.
We find that the solutions interpolate between the two exact analytical solutions to the higher dimensional Einstein-Maxwell theory, which are based on Eguchi-Hanson type I and II geometries. Moreover, we construct the exact cosmological solutions to the theory, and study the properties of the solutions. 

\end{abstract}
\end{titlepage}\onecolumn 
\bigskip 

\section{Introduction}
Exploring the different aspects of gravitational physics is possible through finding the new solutions to gravity, especially coupled to the other fields, such as the electromagnetic field.  
{\textcolor{black}{
Moreover, the possibility of extending the known solutions in asymptotically flat spacetime to the asymptotically de-Sitter and anti de-Sitter solutions, is crucial and important in high energy physics. These extended solutions provide 
better understanding of the holographic proposals between the extended theories of gravity and the conformal field theories in different dimensions \cite{new1}-\cite{new2}. The constructed and explored solutions also include different solutions with different charges, such as NUT charges \cite{new3}, dyonic charges \cite{new4}, as well as different matter fields \cite{new5}, 
axion field \cite{new6} and skyrmions \cite{new7}. The theories of gravity coupled to the other matter fields are useful to study and explore the rotating black holes \cite{new8}, topological charged hairy black holes \cite{new9}, cosmic censorship \cite{new10}, gravitational radiation \cite{new11} and hyper-scaling violation \cite{new12}.}} 
Moreover, finding new solutions to the higher dimensional gravitational theories reveals new phenomena and possibilities, which may not exist in four dimensions. The discovery of the black hole solutions in five dimensions with squashed 3-sphere horizon \cite{one}, the black rings with $S^2 \times S^1$ horizon \cite{two} and black lenses \cite{lens} are just some of the rich variety of the black objects in five and higher dimensions.  Other relevant solutions in higher dimensional gravity coupled to the matter fields, are the dyonic solutions \cite{EY2}, the solitonic solutions \cite{X135}, supergravity solutions \cite{other,hashi}, braneworld cosmologies \cite{eight}, and string theory extended solutions \cite {nine,other13}. 

{\textcolor{black}{ The black hole solutions with different type of topologies for the horizon also were constructed and explored in \cite{NEW9}. Moreover, the references \cite{NEW10,NEW11} include the other solutions to extended theories of gravity with different type of matter fields in different dimensions. The class of solutions to Einstein-Maxwell-dilaton theory, in which the dilaton field couples to the cosmological constant and the Maxwell field, was considered in \cite{NEW12}. These solutions are relevant to the generalization of the Freund-Rubin compactification of M-theory \cite{NEW13}.}}
{\textcolor{black}{Moreover, in interesting papers \cite{NEW14}, the authors constructed and explored the convolution-like solutions for the fully localized type IIA D2 branes intersecting with the D6 branes. The type IIA solutions are obtained by compactifying the corresponding convolution-like M2 brane solutions, over a circle of transverse self-dual geometries including the Bianchi type IX geometry. 
The solutions consist of analytically convolution-like integrals of two functions, which depend on the transverse directions to the branes.
The solutions preserve eight supersymmetries and are valid everywhere; near and far from the core of D6 branes. 
Due to the self-duality of the transverse geometries in the constructed solutions, the compactified solutions are supersymmetric.
We also mention that one interesting feature of the solutions is that, the solutions are expressed completely in terms of convolution integrals, that is a result of taking special ansatzes for the solutions, and separability of the field equations. 
}}

{\textcolor{black}{The motivations for this article come from several works on finding exact solutions in different theories of gravity.
Inspired by the convolution-like solutions in M-theory \cite{NEW14}, in this article, we construct similar convolution-like solutions in six and higher dimensional Einstein-Maxwell theory based on Bianchi type IX geometry.
According to our knowledge, there are not any known convolution-like solutions in six dimensional or higher dimensional Einstein-Maxwell theories based on the Binachi geometries.  Moreover, we are inspired with the papers  \cite{NEW15}, in which the authors constructed charged black hole and black string solutions in five dimensional Einstein-Maxwell theory. The solutions are based on embedding some known four-dimensional geometries, like Kasner space into five dimensions by using appropriate ansatzes for the metric and the Maxwell field. We note that to have non-trivial convolution-like solutions, the minimal dimensionality of Einstein-Maxwell theory should be six. Moreover, we consider the Einstein-Maxwell theories with positive cosmological constant in six and higher dimensions and find the exact cosmological convoluted solutions.}}

To find the new solutions to the higher-dimensional Einstein-Maxwell theory, we consider the Bianchi type IX space, which is an exact solution to the four-dimensional Einstein's equations. Different types of solutions to the Einstein-Maxwell theory were constructed, such as solutions with the NUT charge \cite{new}, solitonic and dyonic solutions \cite{EY2}, as well as braneworld solutions \cite{eight}. Moreover, solutions to the extension of Einstein-Maxwell theory with the axion field and Chern-Simons term, were constructed and studied extensively in \cite{neww}.  

We organize the article as follows: In section \ref{sec:BIX}, we consider the physics of Bianchi type IX spaces. In section \ref{sec:6D}, we present some numerical solutions to the Einstein-Maxwell theory in six and higher dimensions, where the metric function can be written as the convolution-like integral of two functions. In section \ref{sec:6Dsecond}, we present the second class of numerical solutions to the Einstein-Maxwell theory in six and higher dimensions. The second class of solutions is completely independent of the solutions in section \ref{sec:6D}. In section \ref{sec:cosmo}, we use the results of sections \ref{sec:6D} and \ref{sec:6Dsecond}, and explicitly construct some cosmological solutions to the Einstein-Maxwell theory with positive cosmological constant, in six and higher dimensions. In section \ref{sec:kok1}, we consider the Bianchi type IX space with the special cases of the Bianchi parameter as $k=0$ and $k=1$. We construct some exact solutions to the Einstein-Maxwell theory in six and higher dimensions, where the radial function involves the Heun-C functions. We discuss the physical properties of the solutions. {\textcolor{black}{We wrap up the article by the concluding remarks and four appendices in section \ref{sec:con}.}}

\section{The Bianchi geometries}
\label{sec:BIX}
The  classification of the homogeneous and isotropic spaces, is crucial to understand the cosmological models of the universe, as well as finding the theoretical models, which are consistent with the experimental data. The first classification of the homogeneous spaces  was done long time ago by Bianchi \cite{BI}.  We know now that there are {\textcolor{black}{eleven}} different homogeneous spaces \cite{Lan}, called  Bianchi type I$,\,\cdots ,$ {\textcolor{black}{VI (class A or B), VII (class A or B), VIII and}}  IX. The Bianchi type IX has been used mainly in cosmological models \cite{cos}, supergravity theories \cite{sixteen} and  extensions of gravity \cite{17}.

{\textcolor{black}{The four-dimensional Bianchi type IX  metric,  is given locally by the line element \cite{GM}
\be
ds^2=e^{2(A(\eta)+B(\eta)+C(\eta))}d\eta^2+ e^{2A(\eta)}\sigma _1^2+ e^{2B(\eta)}\sigma _2^2++ e^{2C(\eta)}\sigma _3^2
\label{metr},
\ee
where $\sigma_i,\, i=1,2,3$ are three Maurer-Cartan forms, and $A,B$ and $C$ are three functions of the coordinate $\eta$.}} The Maurer-Cartan forms are given by 
\begin{eqnarray}
\sigma_1&=&d\psi + \cos \theta\, d \phi,\\
\sigma_2&=&\cos \psi \, d \theta +\sin \psi \sin \theta \, d \phi,\\
\sigma_3&=&-\sin \psi \, d \theta +\cos \psi \sin \theta \, d \phi,
\end{eqnarray}
in terms of three coordinates $\theta, \, \phi, \, $ and $\psi$ of a unit $S^3$. The  line element (\ref{metr}) has an $SU(2)$ isometry group.  
{\textcolor{black}{The metric satisfies exactly the vacuum Einstein's equations, provided the functions $A(\eta),B(\eta)$ and $C(\eta)$ satisfy 
\begin{eqnarray}
2\frac{d^2 A(\eta)}{d\eta^2}=e^{4A(\eta)}-(e^{2B(\eta)}-e^{2C(\eta)})^2,\label{EA}\\
2\frac{d^2 B(\eta)}{d\eta^2}=e^{4B(\eta)}-(e^{2C(\eta)}-e^{2A(\eta)})^2,\label{EB}\\
2\frac{d^2 C(\eta)}{d\eta^2}=e^{4C(\eta)}-(e^{2A(\eta)}-e^{2B(\eta)})^2,\label{EC}
\end{eqnarray}
as well as
\begin{eqnarray}
4\frac{dA(\eta)}{d\eta}\frac{dB(\eta)}{d\eta}+4\frac{dB(\eta)}{d\eta}\frac{dC(\eta)}{d\eta}+4\frac{dC(\eta)}{d\eta}\frac{dA(\eta)}{d\eta}&=&2(e^{2A(\eta)+2B(\eta)}+e^{2B(\eta)+2C(\eta)}+e^{2C(\eta)+2A(\eta)})\nonumber\\
&-&(e^{4A(\eta)}+e^{4B(\eta)}+e^{4C(\eta)}).\label{EE}
\end{eqnarray}
We should notice equation (\ref{EE}) is the first integral of (\ref{EA})-(\ref{EC}). All Bianchi type IX solutions are self-dual geometries, which leads to the following first order differential equations for the metric functions $A(\eta),B(\eta)$ and $C(\eta)$
\begin{eqnarray}
2\frac{d A(\eta)}{d\eta}=-e^{2A(\eta)}+e^{2B(\eta)}+e^{2C(\eta)}-\beta_1e^{B(\eta)+C(\eta)},\label{EEin1}\\
2\frac{d B(\eta)}{d\eta}=-e^{2B(\eta)}+e^{2C(\eta)}+e^{2A(\eta)}-\beta_2e^{C(\eta)+A(\eta)},\label{EEin2}\\
2\frac{d C(\eta)}{d\eta}=-e^{2C(\eta)}+e^{2A(\eta)}+e^{2B(\eta)}-\beta_3e^{A(\eta)+B(\eta)},\label{EEin3}
\end{eqnarray}
where $\beta_i,\,i=1,2,3$ are three integration constants, which satisfy $\beta_i^2=0$ or $4$ and
\begin{equation}
\beta_i\beta_j=2\epsilon _{ijk}\beta_k.\label{betas}
\end{equation}
The solutions to equation (\ref{betas}) are given by
$(\beta _{1},\beta _{2},\beta _{3})=(0,0,0),(\beta _{1},\beta _{2},\beta _{3})=(2,2,2),$
$(\beta _{1},\beta _{2},\beta _{3})=(2,-2,-2),
(\beta _{1},\beta _{2},\beta _{3})=(-2,2,-2),
(\beta _{1},\beta _{2},\beta _{3})=(-2,-2,2)$.
In appendix \ref{app.AH}, we show that we can not construct the exact solutions to the higher dimensional Einstein-Maxwell theory, based on embedding the four dimensional solutions where $(\beta _{1},\beta _{2},\beta _{3})=(2,2,2)$. We also show that the other three cases $(\beta _{1},\beta _{2},\beta _{3})=(2,-2,-2),
(\beta _{1},\beta _{2},\beta _{3})=(-2,2,-2),
(\beta _{1},\beta _{2},\beta _{3})=(-2,-2,2)$ are equivalent to $(\beta _{1},\beta _{2},\beta _{3})=(2,2,2)$. }}

{\textcolor{black}{Hence, the only viable and interesting solution which we consider is $(\beta _{1},\beta _{2},\beta _{3})=(0,0,0)$. In fact, we can solve the set of differential equations (\ref{EEin1})-(\ref{EEin3}) exactly, where $(\beta _{1},\beta _{2},\beta _{3})=(0,0,0)$. We find
the solutions as
\begin{eqnarray}
e^{2A(\eta)}&=& {c^{2}
\frac{\mathfrak{cn}(c^2\eta ,k^2)\mathfrak{dn}(c^2\eta ,k^2)}{\mathfrak{sn}(-c^2\eta ,k^2)}}
, \label{A1} \\
e^{2B(\eta)}&=& {c^{2}
\frac{\mathfrak{cn}(c^2\eta ,k^2)}{\mathfrak{dn}(c^2\eta ,k^2)\mathfrak{sn}(-c^2\eta ,k^2)}}
,\label{A2} \\
e^{2C(\eta)}&=& {c^2
\frac{\mathfrak{dn}(c^2\eta ,k^2)}{\mathfrak{cn}(c^2\eta ,k^2)\mathfrak{sn}(-c^2\eta ,k^2)}}
,\label{A3} 
\end{eqnarray}
where $c$ and $k$ are integration constants, and $\mathfrak{sn}(z,l)$, $\mathfrak{cn}(z,l)$ and $\mathfrak{dn}(z,l)$ are the Jacobi elliptic functions with the variable $z$ and the parameter $0 \leq l \leq 1$. For completeness, we present the explicit forms of the Jacobi elliptic functions $\mathfrak{sn}(z,l)$, $\mathfrak{cn}(z,l)$ and $\mathfrak{dn}(z,l)$ in appendix \ref{app.elliptic}.  By a straightforward calculation, we also find that the exact solutions (\ref{A1})-(\ref{A3}), indeed satisfy the other field equations (\ref{EA})-(\ref{EE}). 
We change the 
coordinate $\eta $ in the metric (\ref{metr})
to the new coordinate $r$, which is given by%
\begin{equation}
r=\frac{2c}{\sqrt{\mathfrak{sn}(c^2\eta ,k^2)}}. \label{rzeta}
\end{equation}%
For simplicity, we choose coordinate $\eta$ in
the range $[0,\alpha_{(c)(k)(1)}]$ where $\alpha_{(c)(k)(m)}$ is the m-th positive root of 
${\mathfrak{sn}(c^2\eta ,k^2)}$. We can equivalently consider any other range of the form $[\alpha_{(c)(k)(2n)},\alpha_{(c)(k)(2n+1)}]$ with 
$n=1,2,3,\cdots$
or
$[-\alpha_{(c)(k)(2n)},-\alpha_{(c)(k)(2n-1)}]$ for the coordinate $\eta$.
In figure \ref{Figreta}, we show the typical behaviour of the new coordinate $r$ versus $\eta$, where $\eta \in [0,\alpha_{(1)(\frac{1}{2})(1)}]$. Note that $\alpha_{(1)(\frac{1}{2})(1)} \simeq 3.193$.
 \begin{figure}[H]
\centering
\includegraphics[width=0.4\textwidth]{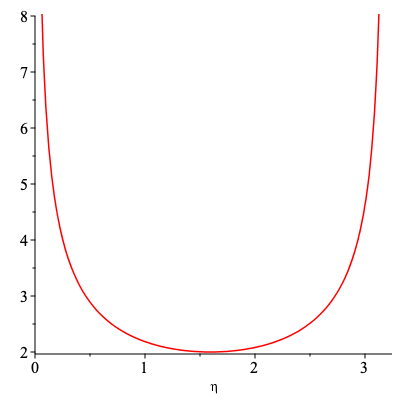}
\caption{The radial coordinate $r$ versus $\eta$, where we set $c=$ and $k=1/2$. The coordinate $\eta \in [0,\alpha_{(1)(\frac{1}{2})(1)}]$, where  $\alpha_{(1)(\frac{1}{2})(1)} \simeq 3.193$.}
\label{Figreta}
\end{figure}
After the change of coordinate (\ref{rzeta}), we find the metric (\ref{metr}) changes to the triaxial Bianchi type IX form, which is given by  \cite{triax}}}
\be
ds^2_{BIX}=\frac{dr^2}{\sqrt{F(r)}}+\frac{r^2\sqrt{F(r)}}{4}(\frac{\sigma_1^2}{1-\frac{a_1^4}{r^4}}+\frac{\sigma_2^2}{1-\frac{a_2^4}{r^4}}+\frac{\sigma_3^2}{1-\frac{a_3^4}{r^4}}),\label{BIX}
\ee
where $a_i,\,i=1,2,3$ are three integration constants, and the metric function $F(r)$ is given by
\be
F(r)=\prod _{i=1}^3 (1-\frac{a_i^4}{r^4}).\label{FF}
\ee
{\textcolor{black}{We note that the three integration constants in (\ref{BIX}), are given by $a_1=0,\, a_2=2kc$ and $a_3=2c$, in terms of the variables and the parameters of the Jacobi elliptic functions $0 \leq k \leq 1$ and $c>0$. We also note that $a_1 \leq a_2 \leq a_3$. }}
The metric (\ref{BIX}) is regular for all values of the radial coordinate $r > 2c$. The Ricci scalar for the Bianchi type IX space (\ref{BIX}) is zero and the Kretschmann invariant is given by
\begin{eqnarray}
{\cal K}&=&\,\frac {1649267441664{c}^{8}}{ \left( 2\,ck-r \right) ^{3} \left( 8
\,{c}^{3}{k}^{3}+4\,{c}^{2}{k}^{2}r+2\,ck{r}^{2}+{r}^{3} \right) ^{3}
 \left( 4\,{c}^{2}+{r}^{2} \right) ^{3} \left( 4\,{c}^{2}-{r}^{2}
 \right) ^{3}{r}^{12}} \nn\\
 &\times&\left(  \left( {\frac {{k}^{8}-k^4+1}{16777216}} \right) {r}^{24}-{
\frac {3\,{c}^{4}{k}^{4} \left( {k}^{4}+1 \right) {r}^{20}}{1048576}}+
{\frac {15\,{c}^{8}{k}^{8}{r}^{16}}{65536}}-{\frac {5\,{c}^{12}{k}^{8}
 \left( {k}^{4}+1 \right) {r}^{12}}{2048}}\right.\nn\\
 &+&\left.{\frac {3\,{c}^{16}{k}^{8}
 \left( {k}^{8}+3\,{k}^{4}+1 \right) {r}^{8}}{256}}-\frac{3}{16}\,{c}^{20}{k}^
{12} \left( {k}^{4}+1 \right) {r}^{4}+{c}^{24}{k}^{16} \right).
\label{KR}
\end{eqnarray}
We notice the Kretschmann invariant (\ref{KR}) is regular everywhere, since $r > 2c$. Moreover, all components of the Ricci tensor are regular, too.

\section{Embedding the Bianchi type IX space in $D\geq 6$-dimensional Einstein-Maxwell theory}
\label{sec:6D}

We consider the $D$-dimensional Einstein-Maxwell theory
\be
S=\int d^Dx \sqrt{-g} (R-\frac{1}{4}F^2),\label{action}
\ee
where $F_{\mu \nu}=\partial _\nu A_\mu -\partial _\mu A_\nu$. {\textcolor{black}{We consider the $D$-dimensional ansatz for the metric, as in \cite{LO}
\begin{equation}
ds_D^{2}=-\frac{dt^{2}}{H_D(r,x)^{2}}+H_D(r,x)^{\frac{2}{D-3}}(dx^2+x^2d\Omega_{D-6}^2+ds^2_{BIX}),
\label{ds6}
\end{equation}
where $ds^2_{BIX}$ is given by (\ref{BIX}) and $d\Omega_{D-6}^2$ is the metric on a unit sphere $S^{D-6}$, where $D \geq 6$. 
We take the components of the  $F_{\mu \nu}$ as in \cite{LO}
\begin{eqnarray}
F_{tr}&=& - \frac{\alpha}{H_D(r,x)^2}\frac{\partial H_D(r,x)}{\partial r} ,\label{gauge1}   \\  F_{tx}&=&- \frac{\alpha}{H_D(r,x)^2}\frac{\partial H_D(r,x)}{\partial x} 
\label{gauge2},
\end{eqnarray}
where $\alpha$ is a constant. We note that (\ref{gauge1}) and (\ref{gauge2}) correspond to the potential $A_t=\frac{\alpha}{H_D(r,x)}$, where all the other components are zero, $A_{\mu \neq t}=0$.}}

\begin{figure}[H]
\centering
\includegraphics[width=0.4\textwidth]{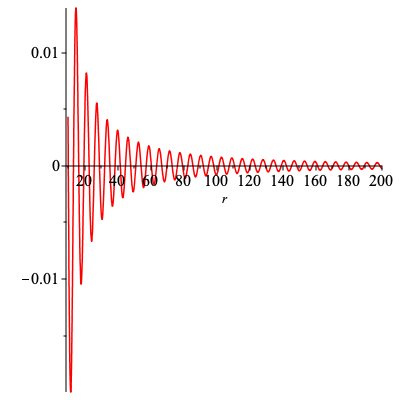}\includegraphics[width=0.4\textwidth]{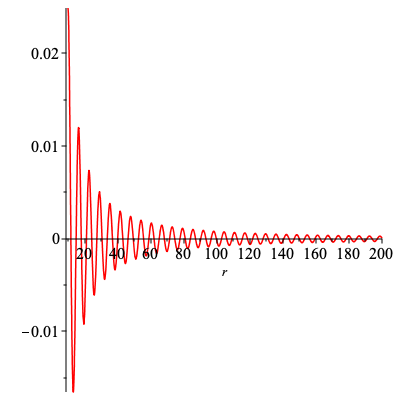}
\caption{The radial function $R(r)$ for large values of $r$, where we set $r_1=1,\,r_2=0$ (left), and $r_1=0,\,r_2=1$ (right) and $g=1$.}
\label{Fig11}
\end{figure}

\begin{figure}[H]
\centering
\includegraphics[width=0.4\textwidth]{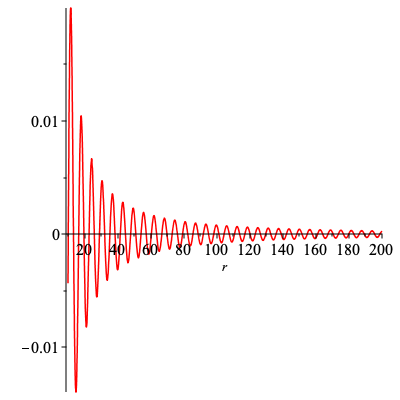}\includegraphics[width=0.4\textwidth]{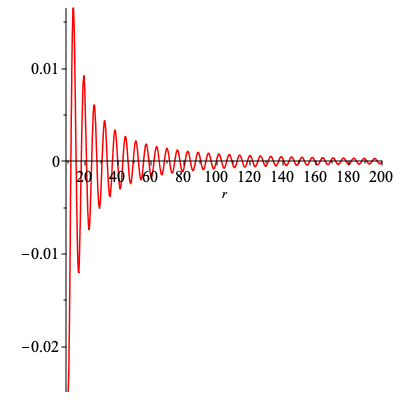}
\caption{The radial function $R(r)$ for large values of $r$, where we set $r_1=1,\,r_2=0$ (left), and $r_1=0,\,r_2=1$ (right) and $g=-1$.}
\label{Figm11}
\end{figure}

{\textcolor{black}{We show in appendix \ref{app:EH}  that all the Einstein's and Maxwell's field equations are satisfied, if the metric function $H(r,x)$ obeys the partial differential equation
\begin{eqnarray}
 &&\left( {\frac {{r}^{9}}{256}}-\frac{1}{16}\,{c}^{4} \left( {k}^{4}+1 \right) 
{r}^{5}+{c}^{8}{k}^{4}r \right)  \left( F \left( r \right) {\frac {
\partial ^{2}}{\partial {r}^{2}}}H_D \left( r,x \right) +\sqrt {F
 \left( r \right) }{\frac {\partial ^{2}}{\partial {x}^{2}}}H_D \left( r
,x \right) + F' \left( r \right) 
  {\frac {\partial }{\partial r}}H_D \left( r,x \right)  \right) \nn\\
&+&7\,F \left( r \right)  \left( {\frac {3\,{r}^{8}}{1792}}-{\frac {5\,{
c}^{4} \left( {k}^{4}+1 \right) {r}^{4}}{112}}+{c}^{8}{k}^{4} \right) 
{\frac {\partial }{\partial r}}H_D \left( r,x \right) =0,\label{Mastereq}
\end{eqnarray}
and the constant $\alpha$ in (\ref{gauge1}) and (\ref{gauge2}) is given by $\alpha^2=\frac{D-2}{D-3}$.}}
To solve the partial differential equation (\ref{Mastereq}), we consider
\be
H_D(r,x)=1+ \beta R(r)X(x),\label{sep}
\ee
where two functions $R(r)$ and $X(x)$ describe the separation of coordinates, and $\beta$ is a constant.  Plugging the equation
(\ref{sep}) into equation (\ref{Mastereq}), we find two ordinary differential equations for the functions $R(r)$ and $X(x)$. The differential equation for the function $X(x)$ is 
\be
x\frac{d^2}{dx^2}X(x)+(D-6)\frac{d}{dx}X(x)-g^2xX(x)=0,\label{Xeq}
\ee
where $g$ denote the separation constant. We find the solutions to  (\ref{Xeq}) are given  by
\be
X(x)=x_1 \frac{I_N (gx)}{x^N}+x_2 \frac{K_N (gx)}{x^N},\label{Xsol}
\ee
where $I_N$ and $K_N$ are the modified Bessel functions of the first and second kind, respectively, and $x_1$ and $x_2$ are the integration constants and 
\be
N=\frac{D-7}{2}.\label{N}
\ee
Moreover, we find the differential equation for the function $R(r)$ as
\begin{eqnarray}
&&\left(-256\,{c}^{8}{k}^{4}+16\,{c}^{4}{k}^{4}{r}^{4}+16\,{c}^{4}{r}
^{4}-{r}^{8} \right)r {\frac {{\rm d}^{2}}{{\rm d}{r}^{2}}}R \left( r
 \right) \nn\\
 &+& \left( 256\,{c}^{8}{k}^{4}+16\,{c}^{4}{k}^{4}{r}^{4}+16\,{c
}^{4}{r}^{4}-3\,{r}^{8} \right) {\frac {\rm d}{{\rm d}r}}R \left( r
 \right)\nn\\
 & -&{g}^{2}{r}^{5}\sqrt {256\,{c}^{8}{k}^{4}-16\,{c}^{4}{k}^{4}{
r}^{4}-16\,{c}^{4}{r}^{4}+{r}^{8}}R \left( r \right) 
 =0.\label{Req}
\end{eqnarray}
We note that the radial differential equation (\ref{Req}) is independent of the dimension $D$ of the spacetime. Tough we can't find any analytic solutions for the equation (\ref{Req}), however we  try to find the analytic solutions to the differential equation (\ref{Req}) in asymptotic region $r \rightarrow \infty$. In the limit of $r \rightarrow \infty$, the equation (\ref{Req}) reduces to
\be
r{\frac {{\rm d}^{2}}{{\rm d}{r}^{2}}}R(r)+3{\frac {{\rm d}^{}}{{\rm d}{r}^{}}}R(r)+rg^2R(r)=0.\label{Reqinf}
\ee
The exact solutions to equation (\ref{Reqinf}) are given by
\be
R(r)=r_1 \frac{J_1 (gr)}{r}+r_2 \frac{Y_1 (gr)}{r},\label{asymR}
\ee
for  $r\rightarrow \infty$, where $J_1$ and $Y_1$ are the Bessel functions of the first and second kind, respectively. 
{\textcolor{black}{In figure \ref{Fig11} and \ref{Figm11}, we plot the behaviour of the $R(r)$ where $r\rightarrow \infty$. We notice the asymptotic radial function monotonically and periodically approaches zero, as $r \rightarrow \infty$, independent of the sign of $g$.}}

\begin{figure}[H]
\centering
\includegraphics[width=0.4\textwidth]{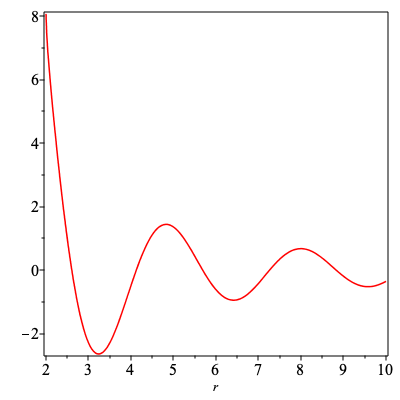}
\caption{The numerical solution for the radial function $R(r)$, where we set $k=\frac{1}{2},\,c=1$ and $g=\pm 2$.}
\label{Fig22}
\end{figure}

Furnished by the asymptotic behaviour of the radial function, we solve numerically the radial differential equation (\ref{Req}) for $0 < k < 1$.  {\textcolor{black}{In figure \ref{Fig22}, we plot the numerical solutions for the radial function $R(r)$, where we set $k=\frac{1}{2},\,c=1$ and $g=\pm 2$.
We should notice that considering $g=- 2$ in the radial differential equation (\ref{Req}) leads to the same radial differential equation with $g=2$. So, we find that the numerical solutions for $g=\pm 2$ are exactly identical, as long as we use the same initial conditions in numerical integration of the differential equation. Hence, in this section, we consider only positive values for the separation constant 
\be 
g\geq 0,\label{gpos}
\ee 
without loosing any generality.}} 
We notice from figure \ref{Fig22} that the radial function becomes divergent as $r\rightarrow 2$, and decays rapidly as $r\rightarrow \infty$, in agreement with the asymptotic solutions (\ref{asymR}) and figures \ref{Fig11} and \ref{Figm11}.

The general structure of the radial function is the same for other values of the Bianchi parameter $c$. The divergent behaviour of the radial function happens at $r \rightarrow 2c$ and the radial function decays rapidly for $r\rightarrow \infty$.

\begin{figure}[H]
\centering
\includegraphics[width=0.4\textwidth]{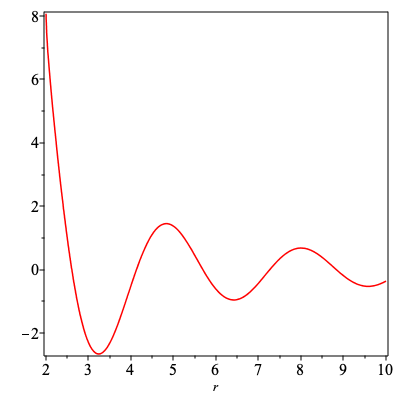}\includegraphics[width=0.4\textwidth]{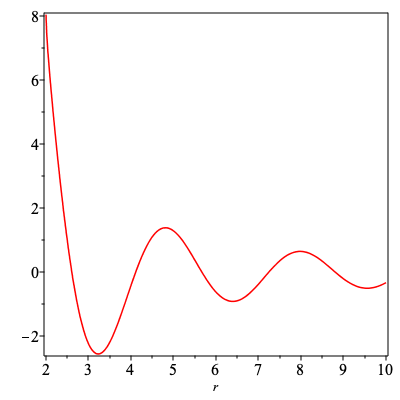}
\caption{The numerical solutions for the radial function $R(r)$, where we set $k=\frac{1}{4}$ (left) and  $k=\frac{3}{4}$ (right) with $c=1,\, g=\pm 2$.}
\label{Fig33}
\end{figure}

Moreover, in figure \ref{Fig33}, we plot the numerical solutions for the radial function $R(r)$, where we set $k=\frac{1}{4}$ and $k=\frac{3}{4}$ with $c=1$ and $g=2$. 

\begin{figure}[H]
\centering
\includegraphics[width=0.4\textwidth]{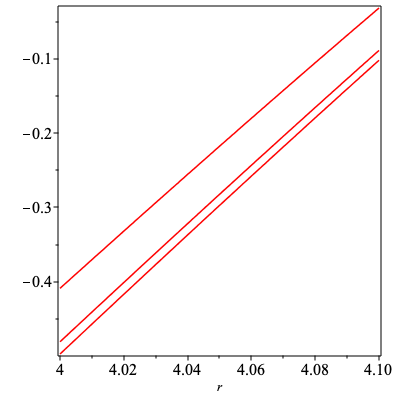}
\caption{The numerical solutions for the radial function $R(r)$, where $k=\frac{3}{4}$ (up),  $k=\frac{1}{2}$ (middle) and $k=\frac{1}{4}$ (down) with $c=1,\, g=2$. }
\label{Fig44}
\end{figure}

\begin{figure}[H]
\centering
\includegraphics[width=0.4\textwidth]{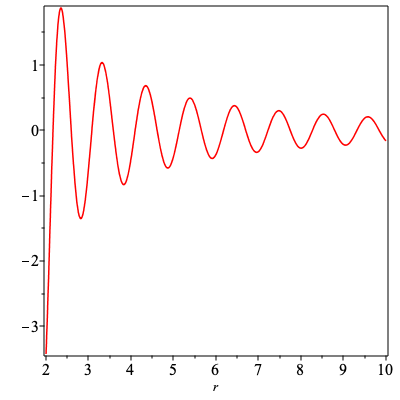}\includegraphics[width=0.4\textwidth]{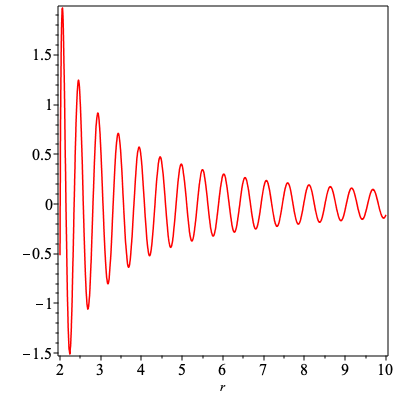}
\caption{The numerical solutions for the radial function $R(r)$, where we set $g=6$ (left) and  $g=12$ (right) with $k=\frac{1}{2},\,c=1$.}
\label{Fig55}
\end{figure}

Tough the figures \ref{Fig22} and \ref{Fig33} are quite similar, however they have subtle dependence on the Bianchi  parameter $k$. In figure \ref{Fig44}, we plot three radial functions, over a small interval of $r$, for $k=\frac{1}{4},\, \frac{1}{2}$ and $\frac{3}{4}$.  As we notice from figure \ref{Fig44}, the radial function $R(r)$, in general, slightly increases with increasing the Bianchi  parameter $k$. 
Changing the separation constant, in general, keeps the overall structure of the radial function. However, increasing the separation constant $g$ leads to more oscillatory behaviour. In figure \ref{Fig55}, we plot the radial functions, for $k=\frac{1}{2}$ and two other separation constants $g=6$ and $g=12$.  

Superimposing all the different solutions with the different  separation constants $g$, we can write the most general solutions to the partial differential equation (\ref{Mastereq}) in $D$-dimensions, as 
\begin{eqnarray}
 H_D(r,x)&=&1+\int _0^\infty \frac{dg}{x^N} \big(P(g) I_N(gx) + Q(g) K_N(gx)\big) R(r),\label{gen}
 \end{eqnarray}
 where $P(g)$ and $Q(g)$ stand for the integration constants, for a specific value of the separation constant $g$, and $N$ is given by (\ref{N}).
 
To find the functions $P(g)$ and $Q(g)$, we may compare the general solutions (\ref{gen}), with another related exact solutions to the theory. In fact, if we consider the large values for the radial coordinate $r \rightarrow \infty$, then the Bianchi type IX metric (\ref{BIX}) changes to
 \be
 d{\cal S}^2=dr^2+\frac{r^2}{4}(d\theta^2+d\phi^2+d\psi^2)+\frac{r^2\cos\theta}{2}d\phi d\psi,\label{asym}
 \ee 
which is the metric on $R^4$. Using the asymptotic metric (\ref{asym}) for the Bianchi type IX geometry, we find an exact solutions to the Einstein-Maxwell theory in $D$-dimensions, where the gravity is described by the metric
 \be
d{\cal S}_D^{2}=-\frac{dt^{2}}{{\cal H}_D(r,x)^{2}}+{\cal H}_D(r,x)^{\frac{2}{D-3}}(dx^2+x^2d\Omega_{D-6}^2+d{\cal S}^2),\label{asymsolD}
 \ee 
 together with the Maxwell's field ${\cal F}_{\mu\nu}$, as
 \begin{eqnarray}
{\cal F}_{tr}&=& - \frac{{\alpha}}{{\cal H}_D(r,x)^2}\frac{\partial {\cal H}_D(r,x)}{\partial r} , \label{ga1}  \\  {\cal F}_{tx}&=&- \frac{\alpha}{{\cal H_D}(r,x)^2}\frac{\partial {\cal H}_D(r,x)}{\partial x} 
\label{gaugeasym}.
\end{eqnarray}
By solving all the Einstein's and Maxwell's field equations, we find that the metric function ${\cal H}_D$ in $D$-dimensions,  is given by the exact form
 \be
 {\cal H}_D(r,x)=1+\frac{\gamma}{(r^2+x^2)^{N+2}},\label{Hasym}
 \ee
where $\gamma$ is a constant and $\alpha^2=\frac{D-2}{D-3}$. {\textcolor{black}{
The $D$-dimensional asymptotic solution (\ref{asymsolD}) with the metric function (\ref{Hasym}) describe a charged spacetime. The Ricci scalar and the Kretschmann invaraint for the solutions  (\ref{asymsolD}) are finite at 
$r=x=0$. The invariants are finite on 
 \be
 {\cal H}_D(r,x)=0,
 \ee
 as long as we consider the constant $\gamma>0$ in (\ref{Hasym}).
 In fact, the Ricci scalar of the solutions (\ref{asymsolD}) at $r=x=0$ is given by
 \be
 {\cal R}_D=\xi_D \frac{(D-3)(D-4)}{\gamma^\frac{2}{D-3}},
 \ee
 where $\xi_6=+1, \, \xi _{D > 6}=-1$. Similarly the Kretschmann invariant of the solutions (\ref{asymsolD}) at $r=x=0$ is given by
 \be
 {\cal K}_D=\frac{\eta_D}{\gamma^\frac{4}{D-3}},
 \ee
 where $\eta_D$ is a constant; $\eta_6=348,\,\eta_7=1064,\,\eta_8=2560,\, \cdots$. The Ricci scalar and the Kretschmann invariant, are given by
 \begin{eqnarray}
  {\cal R}_D&=&\frac{\beta_D\gamma^2}{((r^2+x^2)^{N+2}+\gamma)^\frac{2(D-2)}{D-3}},\\
   {\cal K}_D&=&\frac{f_D(r,x,\gamma)\gamma^2}{((r^2+x^2)^{N+2}+\gamma)^\frac{4(D-2)}{D-3}},
 \end{eqnarray}
 where $\beta_6=6,\,\beta_7=-12,\,\beta_8=-20,\,\cdots$ and some of $f_D(r,x,\gamma)$ are given by
 \begin{eqnarray}
 f_6(r,x,\gamma)&=&12\left( 80\,{r}^{6}+240\,{r}^{4}{x}^{2}+240\,{r}^{2}{x}^{4}+80\,{x
}^{6}-64\,\gamma\,{r}^{2}\sqrt {{r}^{2}+{x}^{2}}-64\,\gamma\,{x}^{2}
\sqrt {{r}^{2}+{x}^{2}}+29\,{\gamma}^{2} \right),\nonumber\\
& &\\
  f_7(r,x,\gamma)&=&8\left( 300\,{r}^{8}+1200\,{r}^{6}{x}^{2}+1800\,{r}^{4}{x}^{4}+1200
\,{r}^{2}{x}^{6}+300\,{x}^{8}-300\,\gamma\,{r}^{4}-600\,\gamma\,{r}^{2
}{x}^{2}-300\,\gamma\,{x}^{4}\right.\nonumber\\
&+&\left.133\,{\gamma}^{2} \right), \\
   f_8(r,x,\gamma)&=&80\left( 63\,{r}^{10}+315\,{r}^{8}{x}^{2}+630\,{r}^{6}{x}^{4}+630\,
{r}^{4}{x}^{6}+315\,{r}^{2}{x}^{8}+63\,{x}^{10}-72\,\gamma\,{r}^{4}
\sqrt {{r}^{2}+{x}^{2}}\right.\nonumber\\
&-&\left.144\,\gamma\,{r}^{2}{x}^{2}\sqrt {{r}^{2}+{x}^
{2}}-72\,\gamma\,{x}^{4}\sqrt {{r}^{2}+{x}^{2}}+32\,
 {\gamma}^{2} \right).
 \end{eqnarray}
As we notice, the Ricci scalar and the Kretschmann invariant are finite on $ {\cal H}_D(r,x)=0$, as long as we choose positive values for $\gamma$, where there are no solutions for ${\cal H}_D(r,x)=0$. 
The electric charge of the black hole solutions (\ref{asymsolD}) is given by
\be
{\cal Q}_D=\frac{1}{2}\int _{\partial \Sigma_{D-1}}ds_{\mu\nu}{\cal F}^{\mu\nu},\label{QQ}
\ee
where $\Sigma_{D-1}$  is a $(D-1)$-dimensional spacelike hypersurface and $\partial \Sigma_{D-1}$ is its boundary. We find the components (\ref{ga1}) and (\ref{gaugeasym}) are given by
\begin{eqnarray}
{\cal F}_{tr}&=&2\alpha\gamma(N+2)\,{\frac { \left( {r}^{2}+{x}^{2} \right) ^{N+1} r}{ \left(  \left( {r}^{2}+{x}^{2} \right) ^{N+2}
 +\gamma \right) ^{2}}},\\
 {\cal F}_{tx}&=&2\alpha\gamma(N+2)\,{\frac { \left( {r}^{2}+{x}^{2} \right) ^{N+1}
  x}{ \left(  \left( {r}^{2}+{x}^{2} \right) ^{N+2}
 +\gamma \right) ^{2}}}.
\end{eqnarray}
Calculating the integral in (\ref{QQ}), we find
\be
{\cal Q}_D=\alpha\gamma\Omega_{D-6}.\label{QQ1}
\ee
We note that $\Omega_{D-6}$ in (\ref{QQ1}) is the volume of a unit sphere $S^{D-6}$ which is given by
\be
\Omega_{D-6}=\frac{2(2\pi)^{(D-6)/2}}{(D-5)!!},\label{OMD6}
\ee
where $n!!$ is equal to $1\cdot 3\cdot 5\cdot \cdots \cdot(2k-1)\cdot(2k+1)$ for an odd $n=2k+1$, and is equal to $\frac{2^{(k+1/2)}k!}{\sqrt{\pi}} $ for an even $n=2k$. 
}}
 
Finding the exact solutions (\ref{asymsolD}) with (\ref{Hasym}), enable us to find an integral equation for the functions $P$ and $Q$  in (\ref{gen}).  In fact, we have the integral equation
 \begin{eqnarray}
 \int _0^\infty \frac{dg}{x^N} \big(P(g) I_N(gx) + Q(g) K_N(gx)\big)  \lim _{ r\rightarrow  \infty} R(r) =
\frac{\gamma }{(r^2+x^2)^{N+2}}.\label{inteq2}
 \end{eqnarray}
 However,  we know $\lim _{ r\rightarrow  \infty} R(r)$ is given by (\ref{asymR}). Considering $r_1=\frac{1}{g},\,r_2=0$, we  solve the integral equation (\ref{inteq2}) and find
 \be
 P(g)=0,\,Q(g)=\frac{\gamma}{2^{N+1}\Gamma (N+2)}g^{N+3},\label{PQ1}
 \ee
 where $\Gamma (N+2)$ is the gamma function. Moreover, considering  the other possibility $r_1=0$, does not lead to consistent solutions for the functions $P$ and $Q$ in the integral equation (\ref{inteq2}).
 Summarizing the results, we find the metric function in $D$-dimensions, as
 \be
  H_D(r,x)=1+\frac{\gamma}{2^{N+1}\Gamma (N+2)}\int _0^\infty \frac{dg}{x^N} g^{N+3} K_N(gx) R(r).\label{genfin}
 \ee
 {\textcolor{black}{We should note that the radial function $R(r)$ as a part of integrand in (\ref{genfin}), indeed depends on integration parameter $g$. Though it is not feasible to find explicitly the Ricci scalar and the Kretschmann invariant of the spacetime (\ref{ds6}), as functions of the coordinates (due to the semi-analytic metric function (\ref{genfin})), however, we may expect that the solutions are regular everywhere (outside of an event horizon with a singularity at $r=2c$), as they approach smoothly to their asymptotic limit (\ref{asymsolD}) with the metric function (\ref{Hasym}). We also expect the electric charge of the spacetime (\ref{ds6}) is the same as equation (\ref{QQ1}). As the metric ansatz (\ref{ds6}) and (\ref{gauge1}) and (\ref{gauge2}) are similar to \cite{NEW15}, we also expect the spacetime (\ref{ds6}) can describe the coalescence of the extremal charged black holes in $D\geq 6$-dimensions, where the spatial section of the black holes consists a copy of the four-dimensional Bianchi type IX.  The extremality is coming from noting that the only free parameter in the metric function (\ref{genfin}) is $\gamma$. Hence the total mass of the gravitational system should be a multiple of $\gamma$. If the electric charge of the spacetime is given by (\ref{QQ1}), then we find that the total mass and the electric charge of the solution are proportional to each other. Of course, it is not feasible to find and verify explicitly the extremality of the solutions, due to the  semi-analytic metric function (\ref{genfin}).}}
 \section{The second class of solutions in $D\geq 6$-dimensional Einstein-Maxwell theory}
\label{sec:6Dsecond}

\begin{figure}[H]
\centering
\includegraphics[width=0.4\textwidth]{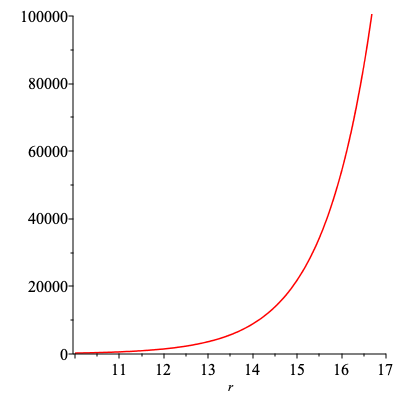}\includegraphics[width=0.4\textwidth]{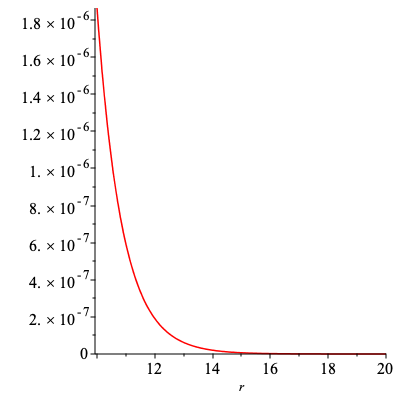}
\caption{The radial function $\tilde R(r)$ for large values of $r$, where we set $\tilde r_1=1,\,\tilde r_2=0$ (left), and $\tilde r_1=0,\,\tilde r_2=1$ (right) and $g=\pm1$.}
\label{Fig11second}
\end{figure}

In this section, we present the second independent class of solutions for the metric function which satisfies the partial differential equation (\ref{Mastereq}).  In separation of the coordinates, we replace $g \rightarrow ig$ in the differential equation (\ref{Xeq}). The solutions of the differential equation are
\be
\tilde X(x)=\tilde x_1 \frac{J_N (gx)}{x^N}+\tilde x_2 \frac{Y_N (gx)}{x^N},\label{xsol2}
\ee
in terms of the Bessel functions, where $\tilde x_1$ and $\tilde x_2$ are constants of integration, and $N$ is given by (\ref{N}). 
Moreover the radial differential equation becomes
\begin{eqnarray}
&&\left(-256\,{c}^{8}{k}^{4}+16\,{c}^{4}{k}^{4}{r}^{4}+16\,{c}^{4}{r}
^{4}-{r}^{8} \right)r {\frac {{\rm d}^{2}}{{\rm d}{r}^{2}}}\tilde R \left( r
 \right) + \left( 256\,{c}^{8}{k}^{4}+16\,{c}^{4}{k}^{4}{r}^{4}+16\,{c
}^{4}{r}^{4}-3\,{r}^{8} \right) {\frac {\rm d}{{\rm d}r}}\tilde R \left( r
 \right)\nn\\
 & +&{g}^{2}{r}^{5}\sqrt {256\,{c}^{8}{k}^{4}-16\,{c}^{4}{k}^{4}{
r}^{4}-16\,{c}^{4}{r}^{4}+{r}^{8}}\tilde R \left( r \right) 
 =0.\label{Req2}
\end{eqnarray}
We present the numerical solutions to equation (\ref{Req2})  for $0 < k < 1$, where its asymptotic form is given by
\be
r{\frac {{\rm d}^{2}}{{\rm d}{r}^{2}}}\tilde R(r)+3{\frac {{\rm d}^{}}{{\rm d}{r}^{}}}\tilde R(r)-rg^2\tilde R(r)=0.\label{Reqinfsecond}
\ee
The exact solutions to equation (\ref{Reqinfsecond}) are given by
\be
\tilde R(r)=\tilde r_1 \frac{I_1 (gr)}{r}+\tilde r_2 \frac{K_1 (gr)}{r},\label{asymRsecond}
\ee
{\textcolor{black}{where $I_1$ and $K_1$ are the modified Bessel functions of the first and second kind, respectively, and $\tilde r_1$ and $\tilde r_2$ are constants}}. In figure \ref{Fig11second}, we plot the behaviour of the $\tilde R(r)$ where $r\rightarrow \infty$. We notice two very different behaviours for the asymptotic radial function, as $r \rightarrow \infty$. 
{\textcolor{black}{Moreover, the plots for $\tilde R(r)$ where $r\rightarrow \infty$, with $g=-1$ are exactly the same as in figure \ref{Fig11second}, since the modified Bessel functions $I_1$ and $K_1$ are even functions of $gr$.}}

\begin{figure}[H]
\centering
\includegraphics[width=0.4\textwidth]{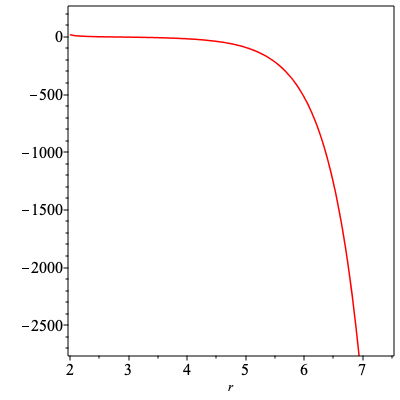}
\caption{The numerical solution for the radial function $\tilde R(r)$, where we set $k=\frac{1}{2},\,c=1$ and $g=\pm 2$.}
\label{Fig22s}
\end{figure}

{\textcolor{black}{ In figure \ref{Fig22s}, we plot the numerical solutions for the radial function $\tilde R(r)$, where we set $k=\frac{1}{2},\,c=1$ and $g=\pm 2$.
We should notice that considering $g=- 2$ in the radial differential equation (\ref{Req2}) leads to the same radial differential equation with $g=2$. So, we find that the numerical solutions for $g=\pm 2$ are exactly identical, as long as we use the same initial conditions in numerical integration of the differential equation. Hence, in this section, we consider only positive values for the separation constant, as in (\ref{gpos}),
without loosing any generality.}} 
We notice from figure \ref{Fig22s} that the radial function approaches zero for $r \rightarrow 2c$, and becomes large, as $r\rightarrow \infty$, in agreement with the asymptotic solutions (\ref{asymRsecond}), where $\tilde r_1=1,\,\tilde r_2=0$, and the left plot in figure \ref{Fig11second}. 

The general structure of the radial function is the same for the other values of the Bianchi parameter $c$. The divergent behaviour of the radial function happens at $r \rightarrow \infty$ and the radial function approaches zero, for $r\rightarrow 2c$.

\begin{figure}[H]
\centering
\includegraphics[width=0.4\textwidth]{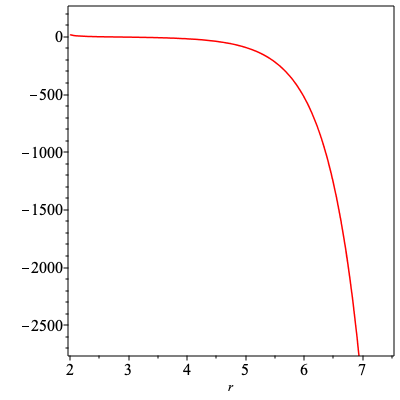}\includegraphics[width=0.4\textwidth]{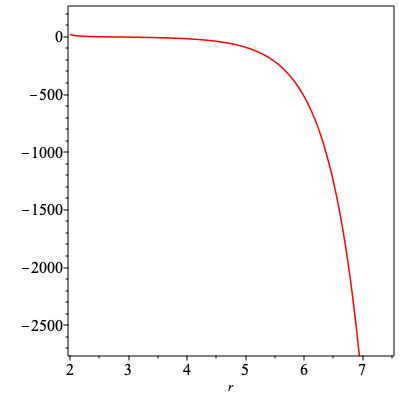}
\caption{The numerical solutions for the radial function $\tilde R(r)$, where we set $k=\frac{1}{4}$ (left) and  $k=\frac{3}{4}$ (right) with $c=1,\, g=2$.}
\label{Fig33s}
\end{figure}

\begin{figure}[H]
\centering
\includegraphics[width=0.4\textwidth]{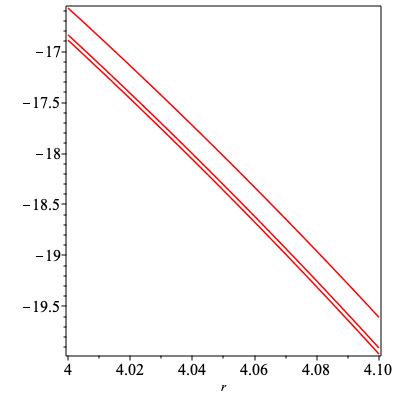}
\caption{The numerical solutions for the radial function $\tilde R(r)$, where $k=\frac{3}{4}$ (up),  $k=\frac{1}{2}$ (middle) and $k=\frac{1}{4}$ (down) with $c=1,\, g=2$. }
\label{Fig44s}
\end{figure}

Moreover, in figure \ref{Fig33s}, we plot the numerical solutions for the radial function $\tilde R(r)$, where we set $k=\frac{1}{4}$ and $k=\frac{3}{4}$ with $c=1$ and $g=2$.  
Tough the figures \ref{Fig22s} and \ref{Fig33s} are quite similar, however they have subtle dependence on the Bianchi  parameter $k$. In figure \ref{Fig44s}, we plot three radial functions, over a small interval of $r$, for $k=\frac{1}{4},\, \frac{1}{2}$ and $\frac{3}{4}$.  As we notice from figure \ref{Fig44s}, the radial function $\tilde R(r)$, in general, slightly increases with increasing the Bianchi  parameter $k$, very similar to the situation for the first radial solutions.

\begin{figure}[H]
\centering
\includegraphics[width=0.4\textwidth]{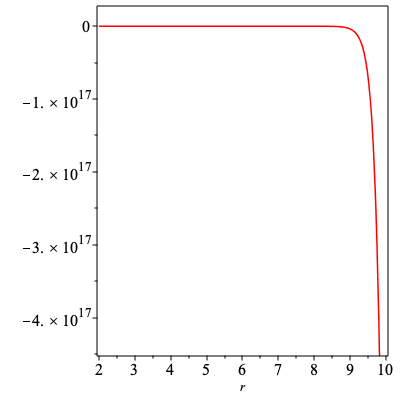}\includegraphics[width=0.4\textwidth]{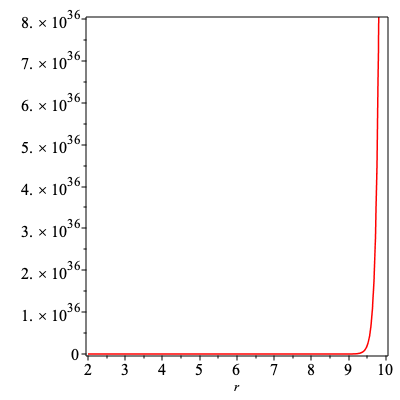}
\caption{The numerical solutions for the radial function $\tilde R(r)$, where we set $g=6$ (left) and  $g=12$ (right) with $k=\frac{1}{2},\,c=1$.}
\label{Fig55s}
\end{figure}

Changing the separation constant, in general, keeps the overall structure of the radial function.
 However, increasing the separation constant $g$ leads to a slower decaying behaviour for the radial function.
 In figure \ref{Fig55s}, we plot the radial functions, for $k=\frac{1}{2}$ and two other separation constants $g=6$ and $g=12$.  
 
 Hence, we find the second most general solutions for the metric function in $D$-dimensions, as given by
\begin{eqnarray}
 \tilde H_D(r,x)&=&1+\int _0^\infty \frac{dg}{x^N} \big(\tilde P(g) J_N(gx) + \tilde Q(g) Y_N(gx)\big) \tilde R(r),\label{gensecond}
 \end{eqnarray}
 where $\tilde P(g)$ and $\tilde Q(g)$ stand for the integration constants, for a specific value of the separation constant $g$ and $N$ is given by (\ref{N}).

 Requiring the metric function (\ref{gensecond}) in the limit of $r\rightarrow \infty$, reduces to the exact solutions (\ref{Hasym}), we find an integral equation for the functions $\tilde P(g)$ and $\tilde Q(g)$, as
 \begin{eqnarray}
 \int _0^\infty \frac{dg}{x^N} \big(\tilde P(g) J_N(gx) + \tilde Q(g) Y_N(gx)\big)  \lim _{ r\rightarrow  \infty} \tilde R(r) =
\frac{\gamma }{(r^2+x^2)^{N+2}}.\label{inteq2sec}
 \end{eqnarray}
 Using equation (\ref{asymRsecond}) for  $\lim _{ r\rightarrow  \infty} \tilde  R(r)$  with  $\tilde r_1=\frac{1}{g},\,\tilde r_2=0$, we  can solve the integral equation (\ref{inteq2sec}) and find
 \be
 \tilde P(g)=\frac{\gamma \Gamma(-N)}{2^{N+2}(N+1)}g^{N+3},\,\tilde Q(g)=0.\label{PQ1}
 \ee
 Moreover, considering  the other possibility $\tilde r_1=0$, does not lead to consistent solutions for the functions $\tilde P$ and $\tilde Q$ in the integral equation (\ref{inteq2sec}).
 Summarizing the results, we find the second metric function in $D$-dimensions, as
 \be
  \tilde H_D(r,x)=1+\frac{\gamma \Gamma(-N)}{2^{N+2}(N+1)}\int _0^\infty \frac{dg}{x^N} g^{N+3} J_N(gx) \tilde R(r).\label{genfinsecond}
 \ee
\section{The solutions in $D\geq 6$ dimensional Einstein-Maxwell theory with cosmological constant}
\label{sec:cosmo}
In this section, we consider the $D\geq 6$ Einstein-Maxwell theory with a cosmological constant $\Lambda$.  We show that there are non-trivial solutions to the $D\geq 6$ dimensional Einstein-Maxwell theory with the cosmological constant. We start by considering the  $D\geq 6$ dimensional metric, as 
\be
ds_{D}^{2}=-H_D(t,r,x)^{-2}dt^{2}+H_D(t,r,x)^{\frac{1}{N+2}}(dx^2+x^2d\Omega_{D-6}+ds_{BIX}^2),\label{metrLambda}
\ee
where the metric function depends on the coordinate $t$, as well as the coordinates $r$ and $x$. The components of the Maxwell's field, are given by equations (\ref{gauge1}) and (\ref{gauge2}), after substitution $H_D(r,x)$ to $H_D(t,r,x)$.  
{\textcolor{black} {We should note considering the ansatzes (\ref{ds6})-(\ref{gauge2}) with metric function $H$ depending on more spatial directions lead to inconsistencies (appendix \ref{app:t}). Inspired with the well-known time-dependent metric functions in the spacetimes with cosmological constant (such as expanding/contracting patches of the de-Sitter spacetime or FLRW spacetime), the above-mentioned inconsistencies could be resolved only by considering a time dependent metric function in the metric, as in (\ref{metrLambda}). Moreover, the components of the Maxwell's field, are given by 
\begin{eqnarray}
F_{tr}&=& - \frac{\alpha}{H_D(t,r,x)^2}\frac{\partial H_D(t,r,x)}{\partial r} ,\label{gauge1T}   \\  F_{tx}&=&- \frac{\alpha}{H_D(t,r,x)^2}\frac{\partial H_D(t,r,x)}{\partial x} 
\label{gauge2T}.
\end{eqnarray}
}} A lengthy calculation shows that all the Einstein's and Maxwell's field equations are satisfied by taking the separation of variables  in the metric function, as
\be
H_D(t,r,x)={T}(t)+{R}(r){X}(x),\label{sep3}
\ee
where the functions $R(r)$ and $X(x)$ satisfy exactly equations (\ref{Req}) and (\ref{Xeq}) for the first class of solutions, respectively, and $\alpha^2=\frac{D-2}{D-3}$. Of course the analytical continuation $g \rightarrow ig$, yields the corresponding equations for $\tilde R(r)$ and $\tilde X(x)$. Moreover, we find 
\be
T(t)=1+\lambda t,
\ee
where 
\be \lambda=(D-3)\sqrt{\frac{2\Lambda}{(D-2)(D-1)}}.\ee
Of course, we should consider only the positive cosmological constant $\Lambda$ to have a real $\lambda$. 
To summarize, we find two classes of the cosmological solutions, where the metric functions are given by
\be
H_D(t,r,x)=1+\lambda t+\frac{\gamma}{2^{N+1}\Gamma (N+2)}\int _0^\infty \frac{dg}{x^N} g^{N+3} K_N(gx) R(r),\label{genfincosmo}
\ee
and
\be
  \tilde H_D(t,r,x)=1+\lambda t+\frac{\gamma \Gamma(-N)}{2^{N+2}(N+1)}\int _0^\infty \frac{dg}{x^N} g^{N+3} J_N(gx) \tilde R(r).\label{genfincosmosecond}
\ee
We note that the two metric functions  (\ref{genfincosmo}) and (\ref{genfincosmosecond}) definitely approach to the exact metric function 
\be
{\cal H}_D(t,r,x)=1+\lambda t+\frac{\gamma}{(r^2+x^2)^{N+2}},\label{Hasymcosmo}
\ee
in the limit of large radial coordinate. The metric function (\ref{Hasymcosmo}) is an exact solution to the Einstein's and Maxwell's field  equations with a cosmological constant $\Lambda$, where 
\be
d{\cal S}_D^{2}=-\frac{dt^{2}}{{\cal H}_D(t,r,x)^{2}}+{\cal H}_D(t,r,x)^{\frac{2}{D-3}}(dx^2+x^2d\Omega_{D-6}^2+d{\cal S}^2),\label{asymsolDcosmo}
 \ee
 and $ d{\cal S}^2$ is given by (\ref{asym}).
 
To get a glimpse of the solution (\ref{metrLambda}), we consider the very late time coordinate, and find 
\be
ds_{D}^{2}=-d\tau^{2}+e^{\frac{\lambda}{N+2}\tau}(dx^2+x^2d\Omega_{D-6}+ds_{BIX}^2),\label{metrLambdatinf}
\ee
where 
\be
\tau=\frac{\ln (1+\lambda t)}{\lambda}\simeq \frac{\ln (\lambda t)}{\lambda}.
\ee
The constant-$\tau$ hypersurface of the metric (\ref{metrLambdatinf}) describes the foliation of $(D-1)$-dimensional de-Sitter space, by flat spacelike Bianchi type IX. The volume of such a hypersurface reaches its minimum at $\tau=0$, and increases exponentially with $\tau$.  {\textcolor{black} {The Hubble parameter ${\bf H}_D$ for the $D$-dimensional metric (\ref{metrLambdatinf}) is equal to 
\be 
{\bf H}_D=\frac{\lambda}{2(N+2)}=\sqrt{\frac{2\Lambda}{(D-2)(D-1)}},\label{HU}
\ee
and the cosmic volume of the spacetime (\ref{metrLambdatinf}) is given by $\int _0^\infty {\cal V}_D(\tau,r,x) d\tau drdx$, where
\be
{\cal V}_D(\tau,r,x)=\frac{1}{4}e^{\frac{(D-1)\lambda}{2(N+2)}\tau}r^3x^{D-6} \Omega_{D-6}.\label{VD}
\ee
We note that $\Omega_{D-6}$ in (\ref{VD}) is the volume of a unit sphere $S^{D-6}$ which is given by equation (\ref{OMD6}).
We also find the cosmic volume of the constant-$\tau$ hypersurface of the metric (\ref{metrLambdatinf}) is  equal to $\int _0^\infty {\cal V}_D(\tau=\text{constant},r,x) drdx$.
As we notice from figure \ref{FigHUB}, the Hubble parameter (\ref{HU}) reduces rapidly with increasing the dimensionality of the spacetime (\ref{metrLambdatinf}).
\begin{figure}[H]
\centering
\includegraphics[width=0.4\textwidth]{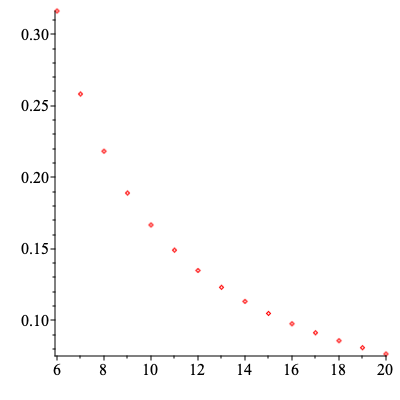}
\caption{The Hubble parameter ${\bf H}_D$ versus the cosmological dimension, where we set $\Lambda=1$.}
\label{FigHUB}
\end{figure}
We note that the metric (\ref{metrLambdatinf})  describes the big bang patch. This patch covers half the dS spacetime from a big bang initiated at past horizon. The bang continues into the Bianchi foliations at future infinity. The other half of dS spacetime is covered by a big crunch patch. This patch initiates at past infinity with the Bianchi foliations. The crush continues towards the future horizon. The contracting patch of solutions at future infinity implies that black hole solutions based on Bianchi type IX space can describe the coalescence of the black holes in asymptotically dS spacetimes.
Moreover, there is a correspondence between the phenomena occurring near the boundary (or in the deep interior) of asymptotically dS (or AdS) spacetime and the ultraviolet (infrared) physics in the dual CFT \cite{DSCFT}. As a result, any solutions in asymptotically (locally) dS spacetimes lead to interpretation in terms of renormalization group flows, and the associated generalized c-theorem. The renormalization group flows toward the infrared in any contracting patch of dS spacetimes. Moreover  the renormalization group flows toward the ultraviolet in any expanding patch of dS spacetimes.  A useful quantity to represent the dS metric with different boundary geometries, such as direct products of flat space, the sphere and hyperbolic space, is the c-function. The c-function for the $D$-dimensional asymptotically dS spacetime (\ref{metrLambdatinf}) is given by \cite{CF}
\be
c\sim\frac{1}{(G_{ij}n^in^j)^{(D-2)/2}},\label{CFUN0}
\ee
where $n^i$ is the unit normal vector to the hypersurface which gives the Hamiltonian constraint and $G$ is the Einstein tensor \cite{CF}. The c-function should show an increasing (decreasing) behaviour versus time for any expanding (contracting) patch of the spacetime.
}}

\begin{figure}[H]
\centering
\includegraphics[width=0.4\textwidth]{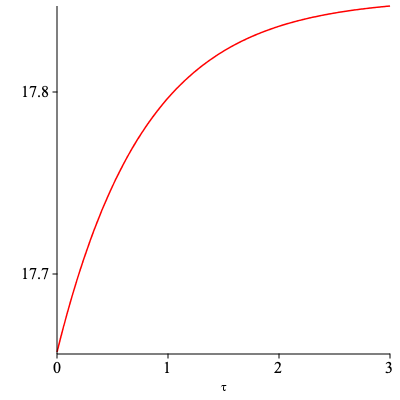}\includegraphics[width=0.4\textwidth]{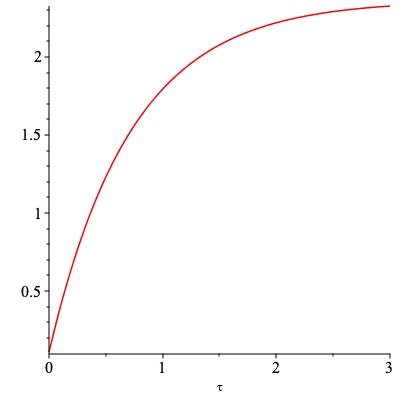}
\caption{The $c$-function of the spacetime (\ref{metrLambdatinf}) versus the cosmological time coordinate $\tau$ for  $D=6$ (left) and $D=7$ (right), where we set $k=\frac{1}{2},\, c=1,\, \Lambda=4,\, r=3$.}
\label{FigCfun}
\end{figure}

In figure \ref{FigCfun}, we  plot  the behaviour of the $c$-function (\ref{CFUN0} ) for two different spacetime dimensions.  Both figures show increasing behaviour for the $c$-function which shows expansion of the constant-$\tau$ hypersurface with the cosmological time $\tau$. Moreover,  we notice that increasing the dimension of spacetime leads to deceasing the value of the $c$-function. So, the constant-$\tau$ hypersurfaces expand slower in the higher dimensions. 

\section{The $D \geq 6$ solutions with $k=0$ and $k=1$ in Einstein-Maxwell theory}
\label{sec:kok1}

In previous sections, we constructed solutions to $D\geq 6$-dimensional Einstein-Maxwell theory, with and without cosmological constant, based on Bianchi type IX geometry with $0 < k <1$. In this section, we consider such solutions, where $k=0$ and $k=1$. 
\subsection{$k=0$}
\begin{figure}[H]
\centering
\includegraphics[width=0.4\textwidth]{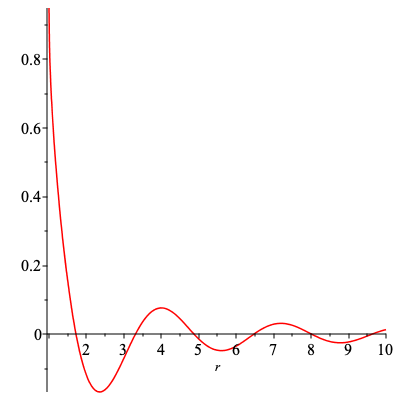}\includegraphics[width=0.4\textwidth]{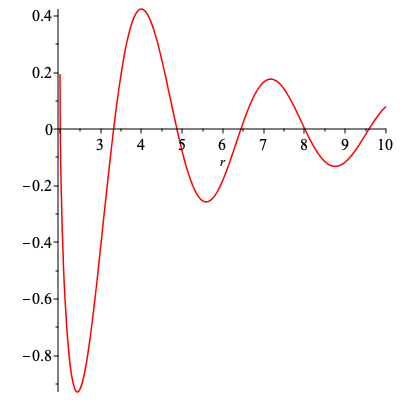}
\caption{The radial function $R(r)$ for $k=0$, where we set $a=1$ (left), and $a=2$ (right) and $g=2$.}
\label{Figk0}
\end{figure}

First we consider the Bianchi type IX metric (\ref{BIX}) with $k=0$.  The metric (\ref{BIX}) reduces to 
\be
ds^2_{k=0}=\frac{dr^2}{\sqrt{1-\frac{a^4}{r^4}}}+     \frac{\sqrt{ r^4-a^4}}{4}        ({\sigma_1^2}+{\sigma_2^2})+
\frac{r^2}{4}\frac{\sigma_3^2}{\sqrt{1-\frac{a^4}{r^4}}},\label{EHI}
\ee
where we set $a=2c$ and the radial coordinate $r \geq a$. The metric (\ref{EHI}) is the metric for the Eguchi-Hanson type I geometry
{\textcolor{black} {\cite{EHIII}.}} The Ricci scalar for (\ref{EHI}) is identically zero, and the Kretschmann invariant is 
\be
{\cal K}=\frac{384 a^8}{(r^2+a^2)^3(r^2-a^2)^3},
\ee
which indicates $r=a$ is a singularity. 
The radial differential equation for $R(r)$ is given by
\be
r(a^4-r^4) {\frac {{\rm d}^{2}}{{\rm d}{r}^{2}}}R \left( r \right) 
 +(a^4-3r^4){\frac {\rm d}{{\rm d}r}}R \left( r
 \right) -{g}^{2}{r}^{3}\sqrt {{r}^{4}-{a}^{4}}R \left( r \right)=0,
\label{Reqk0}
\ee
while the differential equation for $X(x)$ is the same as (\ref{Xeq}) with the solutions (\ref{Xsol}). We find the real analytical solutions for (\ref{Reqk0}), which is
\begin{eqnarray}
R(r)&=&\frac{1}{2}\,{\bold H}_C\big( 0,0,0,\frac{ig^2a^2}{2},-\frac{ig^2a^2}{4},\frac{a^2-i\sqrt{r^4-a^4}}{2a^2}\big)\nn\\
&+&\frac{1}{2}\,{\bold H}^*_C\big( 0,0,0,\frac{ig^2a^2}{2},-\frac{ig^2a^2}{4},\frac{a^2-i\sqrt{r^4-a^4}}{2a^2}\big),
\end{eqnarray}
where $\bold H_C$ is the Heun-C function. In figure \ref{Figk0}, we plot the radial function for different values of the Eguchi-Hanson parameter $a$. We note that the radial function has the oscillatory behaviour similar to cases where $0 < k <1$, however the radial function remains finite at $r=a$.

Combining the different solutions for the radial function $R(r)$ and $X(x)$, we find the most general solution for the metric function  $H_D(r,x)$ in $D$-dimensions, where $k=0$, as 

\begin{eqnarray}
 H_D(r,x)&=&1+\int _0^\infty \frac{dg}{x^N} \big(P_0(g) I_N(gx) + Q_0(g) K_N(gx)\big)\nn\\
&\times&\Re \big({\bold H}_C\big( 0,0,0,\frac{ig^2a^2}{2},-\frac{ig^2a^2}{4},\frac{a^2-i\sqrt{r^4-a^4}}{2a^2}\big)\big),
\label{genk0}
 \end{eqnarray}
 where $P_0(g)$ and $Q_0(g)$ stand for the integration constants, for a specific value of the separation constant $g$.
To find the functions $P_0(g)$ and $Q_0(g)$, we consider the limit $r \rightarrow \infty$, where the Eguchi-Hanson type I space (\ref{EHI}) becomes
\be
ds^2_{k=0}={dr^2}+ \frac{ r^2}{4}({\sigma_1^2}+{\sigma_2^2}+{\sigma_3^2}). \label{EHIinf}
\ee
The asymptotic line element (\ref{EHIinf}) describes $R^4$, and embedding it in the $D$-dimensional theory, leads to the exact solution (\ref{asymsolD})-(\ref{gaugeasym}) with the metric function (\ref{Hasym}). We find the integral equation
\begin{eqnarray}
&&\int _0^\infty \frac{dg}{x^N} \big(P_0(g) I_N(gx) + Q_0(g) K_N(gx)\big)\lim_{r\rightarrow \infty}
\Re \big({\bold H}_C\big( 0,0,0,\frac{ig^2a^2}{2},-\frac{ig^2a^2}{4},\frac{a^2-i\sqrt{r^4-a^4}}{2a^2}\big)\big)\nn\\
&=&\frac{\gamma}{(r^2+x^2)^{N+2}},\label{integk0}
\end{eqnarray}
for the functions $P_0(g)$ and $Q_0(g)$. Moreover, we find
\be
\lim_{r\rightarrow \infty} \big({\bold H}_C\big( 0,0,0,\frac{ig^2a^2}{2},-\frac{ig^2a^2}{4},\frac{a^2-i\sqrt{r^4-a^4}}{2a^2}\big)\big) = \frac{2}{gr}J_1(gr).
\ee
After a very long calculation, we find the solutions to the integral equation (\ref{integk0}), as
\be
P_0(g)=\frac{\gamma \Gamma (-N)}{2^{N+3}(N+1)}g^{N+3},\,Q_0(g)=0.\label{P0Q0k0}
\ee
Hence, we find the exact solutions for the metric function $H(r,x)$ in $D$-dimensional Einstein-Maxwell theory with an embedded four-dimensional Eguchi-Hanson type I geometry (\ref{EHI}) in the spatial section of the spacetime, as
\begin{eqnarray}
 H_D(r,x)&=&1+\frac{\gamma \Gamma (-N)}{2^{N+3}(N+1)}\int _0^\infty \frac{dg}{x^N}g^{N+3} I_N(gx)
\Re \big({\bold H}_C\big( 0,0,0,\frac{ig^2a^2}{2},-\frac{ig^2a^2}{4},\frac{a^2-i\sqrt{r^4-a^4}}{2a^2}\big)\big).\nn\\
&&
\label{genk0fin}
 \end{eqnarray}
We also verify that our numerical solutions to the differential equation (\ref{Req}), where $k=0$, represent exactly the profile of the Heun-C function in equation (\ref{genk0fin}). 

\begin{figure}[H]
\centering
\includegraphics[width=0.4\textwidth]{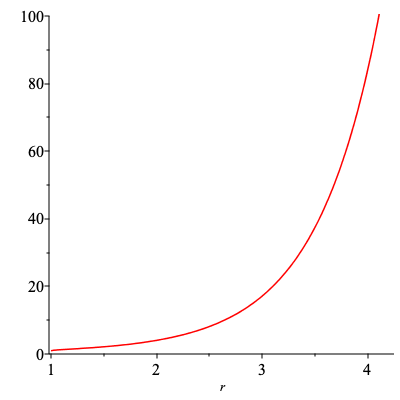}\includegraphics[width=0.4\textwidth]{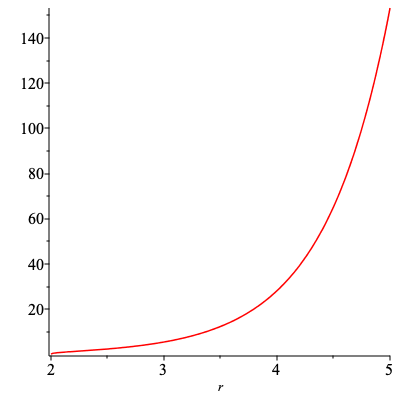}
\caption{The radial function $\tilde R(r)$ for $k=0$, where we set $a=1$ (left), and $a=2$ (right) and $g=2$.}
\label{Figk0second}
\end{figure}

Changing the separation constant $g \rightarrow ig$ generates the second class of solutions, where $k=0$. We find the radial equation 
\be
r(a^4-r^4) {\frac {{\rm d}^{2}}{{\rm d}{r}^{2}}}\tilde R \left( r \right) 
 +(a^4-3r^4){\frac {\rm d}{{\rm d}r}}\tilde R \left( r
 \right) +{g}^{2}{r}^{3}\sqrt {{r}^{4}-{a}^{4}}\tilde R \left( r \right)=0,
\label{tildeReqk0}
\ee
where the exact solutions are given by
\begin{eqnarray}
\tilde R(r)&=&\frac{1}{2}\,{\bold H}_C\big( 0,0,0,-\frac{ig^2a^2}{2},\frac{ig^2a^2}{4},\frac{a^2-i\sqrt{r^4-a^4}}{2a^2}\big)\nn\\
&+&\frac{1}{2}\,{\bold H}^*_C\big( 0,0,0,-\frac{ig^2a^2}{2},\frac{ig^2a^2}{4},\frac{a^2-i\sqrt{r^4-a^4}}{2a^2}\big).
\end{eqnarray}
In figure \ref{Figk0second}, we plot the radial function $\tilde R$ for different values of the Eguchi-Hanson parameter $a$. 

We then superimpose the different solutions for the radial function $\tilde R(r)$ and $\tilde X(x)$, as given by (\ref{xsol2}), to find the second most general solution for the metric function $\tilde H_D(r,x)$ in $D$-dimensions, where $k=0$, as 

\begin{eqnarray}
 \tilde H_D(r,x)&=&1+\int _0^\infty \frac{dg}{x^N} \big(\tilde P_0(g) J_N(gx) + \tilde Q_0(g) Y_N(gx)\big)\nn\\
&\times&\Re \big({\bold H}_C\big( 0,0,0,-\frac{ig^2a^2}{2},\frac{ig^2a^2}{4},\frac{a^2-i\sqrt{r^4-a^4}}{2a^2}\big)\big),
\label{genk0second}
 \end{eqnarray}
 where $\tilde P_0(g)$ and $\tilde Q_0(g)$ stand for the integration constants, for a specific value of the separation constant $g$.
To find the functions $\tilde P_0(g)$ and $\tilde Q_0(g)$, we consider the limit $r \rightarrow \infty$, where the Eguchi-Hanson type I space (\ref{EHI}) becomes (\ref{EHIinf}). Hence we find the integral equation

\begin{eqnarray}
&&\int _0^\infty \frac{dg}{x^N} \big(\tilde P_0(g) J_N(gx) + \tilde Q_0(g) Y_N(gx)\big)\lim_{r\rightarrow \infty}
\Re \big({\bold H}_C\big( 0,0,0,-\frac{ig^2a^2}{2},\frac{ig^2a^2}{4},\frac{a^2-i\sqrt{r^4-a^4}}{2a^2}\big)\big)\nn\\
&=&\frac{\gamma}{(r^2+x^2)^{N+2}},\label{integk0second}
\end{eqnarray}
for the functions $\tilde P_0(g)$ and $\tilde Q_0(g)$.  We find
\be
\lim_{r\rightarrow \infty} \big({\bold H}_C\big( 0,0,0,-\frac{ig^2a^2}{2},\frac{ig^2a^2}{4},\frac{a^2-i\sqrt{r^4-a^4}}{2a^2}\big)\big) = \frac{2}{gr}I_1(gr).
\ee
After a very long calculation, we find the solutions to the integral equation (\ref{integk0second}), as
\be
\tilde P_0(g)=\frac{(-1)^{N+1}\gamma \Gamma (-N-1)}{2^{N+3}}g^{N+3},\,\tilde Q_0(g)=0.\label{P0Q0k0second}
\ee
Hence, we find the second exact solutions for the metric function $\tilde H(r,x)$ in $D$-dimensional Einstein-Maxwell theory with an embedded four-dimensional Eguchi-Hanson type I geometry (\ref{EHI}) in the spatial section of the spacetime, as
\begin{eqnarray}
 \tilde H_D(r,x)&=&1+\frac{(-1)^{N+1}\gamma \Gamma (-N-1)}{2^{N+3}}\int _0^\infty \frac{dg}{x^N}g^{N+3} J_N(gx)\nonumber\\
&\times&\Re \big({\bold H}_C\big( 0,0,0,-\frac{ig^2a^2}{2},\frac{ig^2a^2}{4},\frac{a^2-i\sqrt{r^4-a^4}}{2a^2}\big)\big).
\label{genk0finsecond}
 \end{eqnarray}
We also verify that our numerical solutions to the differential equation (\ref{Req2}), where $k=0$, represent exactly the profile of the Heun-C function in equation (\ref{genk0finsecond}).

\subsection{$k=1$}
In this section, we consider the Bianchi type IX metric (\ref{BIX}) with $k=1$. The metric (\ref{BIX}) reduces to 
\be
ds^2_{k=1}=\frac{dr^2}{{1-\frac{a^4}{r^4}}}+     \frac{{ r^4-a^4}}{4r^2}        {\sigma_1^2} +
\frac{r^2}{4}(\sigma_2^2+\sigma_3^2),\label{EHII}
\ee
which is the known metric for the Eguchi-Hanson type II space, where $r \geq a$ 
{\textcolor{black} {\cite{EHIII}. }}
The Ricci scalar for (\ref{EHII}) is identically zero, and the Kretschmann invariant is given by
\be
{\cal K}=\frac{384 a^8}{r^{12}},
\ee
which is regular and finite, for $r \geq a$. 

The exact solutions for the metric function $H(r,x)$ in $D$-dimensional Einstein-Maxwell theory with an embedded four-dimensional Eguchi-Hanson type II geometry (\ref{EHII}) in the spatial section of the spacetime, is given by \cite{kumarEH}
\begin{eqnarray}
 H_D(r,x)&=&1+\frac{\gamma}{2\xi_D}\int _0^\infty \frac{dg}{x^N}g^{N+3} K_N(gx)
{\bold H}_C\big( 0,0,0,\frac{-g^2a^2}{2},\frac{g^2a^2}{4},\frac{a^2-r^2}{2a^2}\big),\nn\\
&&
\label{genk1fin}
 \end{eqnarray}
where $\xi_{6+2n}=\sqrt{\frac{\pi}{2}}(2n+1)!!,\, n=0,1,2,\cdots$ for even dimensions $D=6+2n$ and $\xi_{7+2n}=(2n+2)!!,\, n=0,1,2,\cdots$ for odd dimensions $D=7+2n$. Moreover, the second class of solutions, is given by the metric function
\begin{eqnarray}
 \tilde H_D(r,x)&=&1+\frac{\gamma\pi(-)^D}{4\xi_D}\int _0^\infty \frac{dg}{x^N}g^{N+3} J_N(gx)
{\bold H}_C\big( 0,0,0,\frac{g^2a^2}{2},-\frac{g^2a^2}{4},\frac{a^2-r^2}{2a^2}\big).\nn\\
&&
\label{genk1finsec}
 \end{eqnarray}
We also verify that our numerical solutions to the differential equation (\ref{Req}), where $k=1$, represent exactly the profile of the Heun-C functions in equations (\ref{genk1fin}) and (\ref{genk1finsec}). {{\textcolor{black} {We should note that the exact semi-analytical results (\ref{genfin}), (\ref{genfinsecond}), (\ref{genfincosmo}), (\ref{genfincosmosecond}), as well as the exact analytical solutions (\ref{genk0fin}) and (\ref{genk0finsecond}) are all novel results. Moreover, the exact solutions for asymptotic metrics (\ref{Hasym}) and (\ref{Hasymcosmo}) are novel results. These results for the metric function $H(r,x)$ furnish new exact solutions to the Einstein-Maxwell theory and its cosmological extension with a continuous parameter $0 < k<  1$ in any dimension $D$ greater than or equal to six. Moreover, we find new exact analytical solutions to the Einstein-Maxwell theory with $k=0$, where the Bianchi space reduces to the Eguchi-Hanson type I space.}}

\section{Concluding Remarks}
\label{sec:con}

We construct a class of exact solutions to the Einstein-Maxwell theory with a continuous parameter $k$. We find the metric function for the solutions in any dimensions $D \geq 6$ uniquely, as a superposition of all radial solutions with their corresponding solutions in the other spatial direction. We solve and present numerical solutions to the radial equation, as we can't find any analytical solutions to the radial field equation. To find the weight functions in the superposition integral, we present another exact solutions to the Einstein-Maxwell theory, such that the superposition integral approaches to the exact metric function of the another solutions, in an appropriate limit. We find complicated integral equations for the weight functions, that we solve and find the unique solutions for the weight functions.  We also consider the positive cosmological constant, and show the field equations for the Einstein-Maxwell theory with positive cosmological constant, can be separated. We find the exact solutions to the field equations and study the properties of the cosmological solutions. 
{{\textcolor{black} {We notice that the exact semi-analytical results (\ref{genfin}), (\ref{genfinsecond}), (\ref{genfincosmo}) and (\ref{genfincosmosecond}) are the novel results in this article. Moreover, the exact analytical solutions (\ref{genk0fin}), (\ref{genk0finsecond}),  (\ref{Hasym}) and (\ref{Hasymcosmo}) are also novel results.}}
We also consider the special case, where the Bianchi type IX parameter $k$ is zero. We show that the Bianchi type IX geometry reduces to a less-known Eguchi-Hanson type I geometry. We find real analytical solutions to the radial equation, in any dimensions, in terms of the Heun-C functions. We also find the weight functions, such that the superimposed solutions reduce to the known exact solutions, in an appropriate limit. {{\textcolor{black} {We notice that the exact analytical results (\ref{genk0fin}) and (\ref{genk0finsecond}) are the novel results in this article. As the ansatzes for the metric and the Maxwell’s field are similar to \cite{NEW15}, we expect the exact solutions can describe the coalescence of the extremal charged black holes in $D\geq 6$-dimensions, where the spatial section of the black holes consists a copy of the four-dimensional Bianchi type IX.}}

\bigskip

{\Large Acknowledgments}

{\textcolor{black}{The author would like to express his sincere gratitude to the anonymous referees for their interesting comments and suggestions, to improve the quality of the article.}} This work was supported by the Natural Sciences and Engineering Research
Council of Canada. 

\appendix

{\textcolor{black} {
\section{Bianchi type IX solutions with $(\beta _{1},\beta _{2},\beta _{3})=(2,2,2)$ }\label{app.AH}
We can find the exact solutions to equations (\ref{EEin1})-(\ref{EEin3}) where $(\beta _{1},\beta _{2},\beta _{3})=(2,2,2)$. In fact, we find the solutions are given by \cite{Hana}
\begin{eqnarray}
e^{2A(\eta )}&=&\frac{2}{\pi }\frac{\vartheta _{2}(i\eta)\vartheta _{3}^{^{\prime
}}(i\eta)\vartheta _{4}^{^{\prime }}(i\eta)}{\vartheta _{2}^{^{\prime }}(i\eta)\vartheta
_{3}(i\eta)\vartheta _{4}(i\eta)},\label{e2A} \\ 
e^{2B(\eta )}&=&\frac{2}{\pi }\frac{\vartheta _{2}^{^{\prime }}(i\eta)\vartheta
_{3}(i\eta)\vartheta _{4}^{^{\prime }}(i\eta)}{\vartheta _{2}(i\eta)\vartheta _{3}^{^{\prime
}}(i\eta)\vartheta _{4}(i\eta)},\label{e2B} \\ 
e^{2C(\eta )}&=&\frac{2}{\pi }\frac{\vartheta _{2}^{^{\prime }}(i\eta)\vartheta
_{3}^{^{\prime }}(i\eta)\vartheta _{4}(i\eta)}{\vartheta _{2}(i\eta)\vartheta _{3}(i\eta)\vartheta
_{4}^{^{\prime }}(i\eta)},\label{e2C}%
\end{eqnarray}%
where the $\vartheta $'s denote the Jacobi Theta functions 
\begin{equation}
\vartheta _{i}(i\eta )=\vartheta _{i}(0\left| i\eta \right. ),  \label{th}
\end{equation}%
and $^\prime=\frac{d}{d\eta}$.
We recall that in general, the 
Jacobi Theta functions $\vartheta_i,\,i=1,2,3,4 $ are given by
\begin{eqnarray}
\vartheta _{1}(\nu \left| \tau \right. ) &=&\vartheta \left[ _{1}^{1}\right]
(\nu \left| \tau \right. ),  \label{thetas} \\
\vartheta _{2}(\nu \left| \tau \right. ) &=&\vartheta \left[ _{0}^{1}\right]
(\nu \left| \tau \right. ),  \label{th1} \\
\vartheta _{3}(\nu \left| \tau \right. ) &=&\vartheta \left[ _{0}^{0}\right]
(\nu \left| \tau \right. ),  \label{th2} \\
\vartheta _{4}(\nu \left| \tau \right. ) &=&\vartheta \left[ _{1}^{0}\right]
(\nu \left| \tau \right. ), \label{th3}
\end{eqnarray}
where
\begin{equation}
\vartheta \left[ _{b}^{a}\right] (\nu \left| \tau \right. )=\sum_{n\in
Z}e^{i\pi (n-\frac{a}{2})\{\tau (n-\frac{a}{2})+2(\nu -\frac{b}{2})\}},
\label{thy}
\end{equation}%
where $a$ and $b$ are two real numbers.
%
The metric (\ref{metr}) with the metric functions (\ref{e2A})-(\ref{e2C}) describes the Atiyah-Hitchin space. 
The other three possible cases $(\beta _{1},\beta _{2},\beta _{3})=(2,-2,-2),(\beta _{1},\beta _{2},\beta _{3})=(-2,2,-2),(\beta _{1},\beta _{2},\beta _{3})=(-2,-2,2)$ are related to  $(\beta _{1},\beta _{2},\beta _{3})=(2,2,2)$. In fact the metric functions for the solutions$(\beta _{1},\beta _{2},\beta _{3})=(2,-2,-2),(\beta _{1},\beta _{2},\beta _{3})=(-2,2,-2)$ and $(\beta _{1},\beta _{2},\beta _{3})=(-2,-2,2)$ are obtained by the replacements $e^{A}$ to $-e^{A}$,  $e^{B}$ to $-e^{B}$, and $e^{C}$ to $-e^{C}$, respectively in (\ref{metr}).
Upon embedding the Atiyah-Hitchin space $ds^2_{AH}$, into the higher-dimensional ansatz (\ref{ds6})
\begin{equation}
ds_D^{2}=-\frac{dt^{2}}{H(r,x)^{2}}+H(r,x)^{\frac{2}{D-3}}(dx^2+x^2d\Omega_{D-6}^2+ds^2_{AH}),
\label{ds6AH}
\end{equation}
we find a differential equation which can not be separated, unlike the separable differential equation (\ref{Mastereq}) for embedding the Bianchi type IX. 
\section{The Jacobi elliptic functions}\label{app.elliptic}
The Jacobi elliptic functions $\mathfrak{sn}(z,l)$, $\mathfrak{cn}(z,l)$ and $\mathfrak{dn}(z,l)$ with the variable $z$ and the parameter $l$, are related to the Jacobi elliptic function $\mathfrak{am}(z,l)$ by%
\begin{eqnarray}
\mathfrak{sn}(z,l)&=&\sin (\mathfrak{am}(z,l)),  \label{SN}\\
\mathfrak{cn}(z,l)&=&\cos (\mathfrak{am}(z,l)),  \label{CN}\\
\mathfrak{dn}(z,l)&=&\sqrt{1-l^2\mathfrak{sn}^2(z,l)}.  \label{DN}
\end{eqnarray}%
The Jacobi elliptic function $\mathfrak{am}(z,l)$ is the inverse of the trigonometric form
of the elliptic integral of the first kind $\mathfrak{f}(\varphi ,l)$. In other words %
\begin{equation}
\mathfrak{am}(\mathfrak{f}(\sin \phi ,l),l)=\phi.   \label{AM}
\end{equation}%
We note that the elliptic integral of the first kind $\mathfrak{f}(\varphi ,k)$ is
given by the following integral
\begin{equation}
\mathfrak{f}(\varphi ,l)=\int_{0}^{\sin ^{-1}(\varphi )}\frac{d\theta }{%
\sqrt{1-l^{2}\sin ^{2}\theta }}.  \label{F}
\end{equation}
\section{The Einstein and Maxwell's field equations}\label{app:EH}
The Einstein's field equations are given by
\be
\epsilon_{\mu\nu}\defeq G_{\mu\nu}-T_{\mu\nu}=0,\label{EEE}
\ee
while the Maxwell's equations are
\be
\varphi^\nu\defeq F^{\mu\nu}_{;\mu}=0,\label{MMM}
\ee
where $G_{\mu\nu}$ is the Einstein tensor,  $F_{\mu\nu}$ is the Maxwell tensor. The $T_{\mu\nu}$ is the energy-momentum tensor for the electromagnetic field, which is given by
\be
T_{\mu\nu}=  F_{\mu}^{\lambda}F_{\nu\lambda}-\frac{1}{4}F^2g_{\mu\nu}.\label{TTT}
\ee
First, we consider $D=6$, in which the metric ansatz (\ref{ds6}) becomes
\begin{equation}
ds^{2}=-\frac{dt^{2}}{H(r,x)^{2}}+H(r,x)^{\frac{2}{3}}(dx^2+ds^2_{BIX}).
\label{ds666}
\end{equation}
The non-zero components of the Maxwell tensor also are given by (\ref{gauge1}) and (\ref{gauge2}). Equation (\ref{MMM}) leads to
\begin{eqnarray}
 \varphi^t&=&\left( {\frac {{r}^{9}}{256}}-\frac{1}{16}\,{c}^{4} \left( {k}^{4}+1 \right) 
{r}^{5}+{c}^{8}{k}^{4}r \right)  \left( F \left( r \right) {\frac {
\partial ^{2}}{\partial {r}^{2}}}H \left( r,x \right) +\sqrt {F
 \left( r \right) }{\frac {\partial ^{2}}{\partial {x}^{2}}}H \left( r
,x \right) + F' \left( r \right) 
  {\frac {\partial }{\partial r}}H \left( r,x \right)  \right) \nn\\
&+&7\,F \left( r \right)  \left( {\frac {3\,{r}^{8}}{1792}}-{\frac {5\,{
c}^{4} \left( {k}^{4}+1 \right) {r}^{4}}{112}}+{c}^{8}{k}^{4} \right) 
{\frac {\partial }{\partial r}}H \left( r,x \right) ,\label{Mastereq1}
\end{eqnarray}
and all the other components of $\varphi^\nu$ are identically zero. Equation (\ref{EEE}) leads to some long expressions for the components of $\epsilon_{\mu\nu}$. For example,
\begin{eqnarray}
\epsilon_{rr}&=&-  \frac {458752}{3\, \left( F \left( r \right)  \right) ^{5/2}H
 \left( r,x \right)  \left( 16\,{k}^{4}{c}^{4}-{r}^{4} \right) ^{2}
 \left( 16\,{c}^{4}-{r}^{4} \right) ^{2}{r}^{2}} \left( 1/7\, \left( {
c}^{2}{k}^{2}+1/4\,{r}^{2} \right)  \left( ck+r/2 \right)  \left( c-r/
2 \right) 
\right.\nonumber\\
&\times&\left. \left(  \left( {\frac {9\,{r}^{9}}{1024}}-{\frac {9\,{c}^{4
} \left( {k}^{4}+1 \right) {r}^{5}}{64}}+9/4\,{c}^{8}{k}^{4}r \right) 
H \left( r,x \right) {\frac {{\rm d}^{2}}{{\rm d}{r}^{2}}}F \left( r
 \right) \right.\right. \nonumber\\
 &+&\left.\left.  {\frac {\rm d}{{\rm d}r}}F \left( r  
 \right)  \left(  \left( {\frac {{r}^{9}}{256}}-1/16\,{c}^{4} \left( {
k}^{4}+1 \right) {r}^{5}+{c}^{8}{k}^{4}r \right) {\frac {\partial }{
\partial r}}H \left( r,x \right) \right.\right.\right.\nonumber\\
&+&\left.\left.\left.{\frac {63\,H \left( r,x \right) }{4
} \left( {\frac {3\,{r}^{8}}{1792}}-{\frac {5\,{c}^{4} \left( {k}^{4}+
1 \right) {r}^{4}}{112}}+{c}^{8}{k}^{4} \right) } \right)  \right) r
 \left( {c}^{2}+1/4\,{r}^{2} \right)  \left( ck-r/2 \right)  \left( c+
r/2 \right)  \left( F \left( r \right)  \right) ^{3/2}\right.\nonumber\\
&+&\left. \left( 1/7\,
 \left( {c}^{2}{k}^{2}+1/4\,{r}^{2} \right) ^{2} \left( ck+r/2
 \right) ^{2} \left( c-r/2 \right) ^{2}{r}^{2} \left( {c}^{2}+1/4\,{r}
^{2} \right) ^{2} \left( ck-r/2 \right) ^{2} \left( c+r/2 \right) ^{2}
{\frac {\partial ^{2}}{\partial {r}^{2}}}H \left( r,x \right) \right.\right.\nonumber\\
&+&\left.\left.
 \left( {c}^{2}{k}^{2}+1/4\,{r}^{2} \right)  \left( c^2k^2-r^2/4 \right) 
 \left( c^2-r^2/4 \right)  \left( {\frac {3\,{r}^{8}}{1792}}-{\frac {5\,{c
}^{4} \left( {k}^{4}+1 \right) {r}^{4}}{112}}+{c}^{8}{k}^{4} \right) r
 \left( {c}^{2}+1/4\,{r}^{2} \right) 
  {\frac {\partial }{\partial r}}H \left( r,x \right)\right.\right.\nonumber\\
&+&\left.\left.{
\frac {36\,{c}^{4}H \left( r,x \right) }{7} \left(  \left( {\frac {{k}
^{4}}{8192}}+{\frac{1}{8192}} \right) {r}^{12}+{\frac {{c}^{4} \left( 
{k}^{8}-4\,{k}^{4}+1 \right) {r}^{8}}{512}}-\frac{1}{32}
{c}^{8}{k}^{4}
 \left( {k}^{4}+1 \right) {r}^{4}+{c}^{12}{k}^{8} \right) } \right) 
 F \left( r \right) ^{5/2}\right.\nonumber\\
 &-&\left.\frac {9\, \left( ck-r/2
 \right) ^{2} \left( {c}^{2}{k}^{2}+1/4\,{r}^{2} \right) ^{2} \left( c
+r/2 \right) ^{2} \left( ck+r/2 \right) ^{2} \left( c-r/2 \right) ^{2}
{r}^{2} \left( {c}^{2}+1/4\,{r}^{2} \right) ^{2}}{56} \right.\nonumber\\
&\times&\left.
\left(  \left( {
\frac {\rm d}{{\rm d}r}}F \left( r \right)  \right) ^{2}H \left( r,x
 \right) \sqrt {F \left( r \right) }
 -{\frac {8\, \left( F \left( r
 \right)  \right) ^{2}{\frac {\partial ^{2}}{\partial {x}^{2}}}H
 \left( r,x \right) }{9}} \right)  \right), \label{err}
\end{eqnarray}
and
\begin{eqnarray}
\epsilon_{\phi\psi}&=&32768\,\frac {\cos \left( \theta \right) }{{r}^{8} \left( 16\,{k}^{4}
{c}^{4}-{r}^{4} \right)  \left( 16\,{c}^{4}-{r}^{4} \right) H \left( r
,x \right) }\nonumber\\
&\times&
 \left(  \left( -{\frac {{r}^{18}}{393216}}+{\frac {{c}^{4
} \left( {k}^{4}+1 \right) {r}^{14}}{24576}}-{\frac {{c}^{8}{k}^{4}{r}
^{10}}{1536}} \right) F \left( r \right) {\frac {\partial ^{2}}{
\partial {r}^{2}}}H \left( r,x \right) \right.\nonumber\\
&+&\left.
\left( -{\frac {{r}^{18}}{
393216}}+{\frac {{c}^{4} \left( {k}^{4}+1 \right) {r}^{14}}{24576}}-{
\frac {{c}^{8}{k}^{4}{r}^{10}}{1536}} \right) \sqrt {F \left( r
 \right) }{\frac {\partial ^{2}}{\partial {x}^{2}}}H \left( r,x
 \right) \right.\nonumber\\
 &+&\left.
\left( -{\frac {{r}^{18}}{524288}}+{\frac {{c}^{4} \left( {
k}^{4}+1 \right) {r}^{14}}{32768}}-{\frac {{c}^{8}{k}^{4}{r}^{10}}{
2048}} \right) H \left( r,x \right) {\frac {{\rm d}^{2}}{{\rm d}{r}^{2
}}}F \left( r \right) \right.\nonumber\\
&-&\left.\frac {{r}^{9}{\frac {\partial }{\partial r}}H
 \left( r,x \right) }{1536} \left(  \left( {\frac {{r}^{9}}{256}}-1/16
\,{c}^{4} \left( {k}^{4}+1 \right) {r}^{5}+{c}^{8}{k}^{4}r \right) {
\frac {\rm d}{{\rm d}r}}F \left( r \right) \right.\right.\nonumber\\
&+&\left.\left.7\,F \left( r \right) 
 \left( {\frac {3\,{r}^{8}}{1792}}-{\frac {5\,{c}^{4} \left( {k}^{4}+1
 \right) {r}^{4}}{112}}+{c}^{8}{k}^{4} \right)  \right) 
 \right.\nonumber\\
 &+&\left.H \left( r,x
 \right)  \left( -{\frac {11\,{r}^{9}{\frac {\rm d}{{\rm d}r}}F
 \left( r \right) }{2048} \left( {\frac {7\,{r}^{8}}{2816}}-{\frac {9
\,{c}^{4} \left( {k}^{4}+1 \right) {r}^{4}}{176}}+{c}^{8}{k}^{4}
 \right) }\right.\right.\nonumber\\
 &-&\left.\left.{\frac {3\,{r}^{8}F \left( r \right) }{256} \left( {\frac {
{r}^{8}}{768}}-1/24\,{c}^{4} \left( {k}^{4}+1 \right) {r}^{4}+{c}^{8}{
k}^{4} \right) }+ \left( {c}^{8}{k}^{4}-1/8\,{c}^{4}{k}^{4}{r}^{4}+{
\frac {{r}^{8}}{256}} \right) \right.\right.\nonumber\\
&\times&\left.\left. \left( {c}^{8}{k}^{4}-1/8\,{c}^{4}{r}^{
4}+{\frac {{r}^{8}}{256}} \right)  \right)  \right) .\label{eps}
\end{eqnarray}
Moreover, the off-diagonal elements of (\ref{EEE}), such as 
\be \epsilon_{r x}=(\alpha^2-\frac{4}{3})\frac{\frac{\partial H(r,x)}{\partial x}\frac{\partial H(r,x)}{\partial r}}{H^2(r,x)},\ee 
lead to $\alpha^2=\frac{4}{3}$.
We algebraically solve equation $\varphi^t=0$ for ${\frac {
\partial ^{2}H \left( r,x \right)}{\partial {r}^{2}}}$, and find it in terms of other derivatives of $H$. Upon substituting the algebraic result for ${\frac {
\partial ^{2}H \left( r,x \right)}{\partial {r}^{2}}}$ in all components of the Einstein equations (\ref{EEE}) (for example (\ref{err}) and (\ref{eps})), we find 
\be
\epsilon_{\mu\nu}=0,
\ee
exactly. We also note that equation $ \varphi^t=0$ is exactly (\ref{Mastereq}). Similar calculations for $D=7,8,\cdots$ show that all the Einstein's equations are exactly satisfied, upon substitution ${\frac {
\partial ^{2}H \left( r,x \right)}{\partial {r}^{2}}}$ which is obtained by solving algebraically the only non-zero component $\varphi^t$ of the Maxwell's field equations (\ref{MMM}). Moreover, we find $\alpha^2=\frac{D-2}{D-3}$. It is worth to note that the differential equation (\ref{Mastereq1}) does not depend on dimensionality of the spacetime. We always find (\ref{Mastereq1}), for the metric ansatz (\ref{ds6}), in all different dimensions $D \geq 6$.
\section{Time-dependent solutions with cosmological constant}\label{app:t}
In this appendix, we show that considering ansatzes (\ref{ds6})-(\ref{gauge2}) with more dependence on the spatial coordinates in $H$, in $D \geq6$ Einstein-Maxwell theory with a cosmological constant $\Lambda$ leads to inconsistencies. We consider $D=6$ with the metric ansatz 
\begin{equation}
ds_6^{2}=-\frac{dt^{2}}{H(r,x,\theta)^{2}}+H(r,x,\theta)^{\frac{2}{3}}(dx^2+ds^2_{BIX}),
\label{ds63coor}
\end{equation}
where the metric function depends on three spatial coordinates $r,x$ and $\theta$. We also consider the components of the  $F_{\mu \nu}$, as 
\begin{eqnarray}
F_{tr}&=& - \frac{\alpha}{H(r,x,\theta)^2}\frac{\partial H(r,x,\theta)}{\partial r} ,\label{gauge13}   \\  
F_{tx}&=&- \frac{\alpha}{H(r,x,\theta)^2}\frac{\partial H(r,x,\theta)}{\partial x},\\
F_{t\theta}&=&- \frac{\alpha}{H(r,x,\theta)^2}\frac{\partial H(r,x,\theta)}{\partial \theta}
\label{gauge23},
\end{eqnarray}
where $\alpha$ is a constant.
The Einstein's field equations are given by
\be
\epsilon_{\mu\nu}\defeq G_{\mu\nu}+\Lambda g_{\mu\nu}-T_{\mu\nu}=0,\label{EEE3}
\ee
while the Maxwell's equations are
\be
\varphi^\nu\defeq F^{\mu\nu}_{;\mu}=0,\label{MMM3}
\ee
where $\Lambda$ is the cosmological constant. We find the only non-zero component of the Maxwell's equations as
\be
\varphi^t={\cal L}_1(r,x,\theta)+\cos^2\psi{\cal L}_2(r,x,\theta)=0,\label{BE}
\ee
where 
\begin{eqnarray}
{\cal L}_1(x,r,\theta)&=&
\left( 4\,{r}^{12}-64\,{c}^{4} \left( {k}^{4}+1 \right) {r}^{8}+1024
\,{c}^{8}{r}^{4}{k}^{4} \right) \sin \left( \theta \right) {\frac {
\partial ^{2}}{\partial {\theta}^{2}}}H \left( r,x,\theta \right)\nonumber\\
&+&
 \left( {r}^{14}-16\,{c}^{4} \left( {k}^{4}+1 \right) {r}^{10}+256\,{c
}^{8}{k}^{4}{r}^{6} \right) F \left( r \right) \sin \left( \theta
 \right) {\frac {\partial ^{2}}{\partial {r}^{2}}}H \left( r,x,\theta
 \right) \nonumber\\
 &+& \left( {r}^{14}-16\,{c}^{4} \left( {k}^{4}+1 \right) {r}^{
10}+256\,{c}^{8}{k}^{4}{r}^{6} \right) \sin \left( \theta \right) 
\sqrt {F \left( r \right) }{\frac {\partial ^{2}}{\partial {x}^{2}}}H
 \left( r,x,\theta \right)\nonumber\\
 &+&256\, \left( {\frac {\partial }{\partial r
}}H \left( r,x,\theta \right)  \right) {r}^{5}\sin \left( \theta
 \right)  \left(  \left( {\frac {{r}^{9}}{256}}-1/16\,{c}^{4} \left( {
k}^{4}+1 \right) {r}^{5}+{c}^{8}{k}^{4}r \right) {\frac {\rm d}{
{\rm d}r}}F \left( r \right) \right.\nonumber\\
&+&7\left. \left( {\frac {3\,{r}^{8}}{1792}}-{
\frac {5\,{c}^{4} \left( {k}^{4}+1 \right) {r}^{4}}{112}}+{c}^{8}{k}^{
4} \right) F \left( r \right)  \right)\nonumber\\
&-&16384
 \left( {c}^{2}+1/4\,{r}^{2} \right)  \left( {\frac {\partial }{
\partial \theta}}H \left( r,x,\theta \right)  \right) \left( {c}^{2}{k}^{2}+1/4\,{r}^{2} \right) ^{2} \left( c^2
k^2-r^2/4 \right) ^{2} \left( c^2-r^2/4 \right) \cos \left( \theta \right),\nonumber\\
&&
\end{eqnarray}
and
\begin{eqnarray}
{\cal L}_2(x,r,\theta)&=&16384\,\cos \left( \theta \right)  \left( {\frac {\partial }{\partial 
\theta}}H \left( r,x,\theta \right)  \right) {c}^{12}{k}^{8}-16384\,
\sin \left( \theta \right)  \left( {\frac {\partial ^{2}}{\partial {
\theta}^{2}}}H \left( r,x,\theta \right)  \right) {c}^{12}{k}^{8}\nn\\
&-&1024
\,\cos \left( \theta \right)  \left( {\frac {\partial }{\partial 
\theta}}H \left( r,x,\theta \right)  \right) {c}^{8}{k}^{4}{r}^{4}-
1024\,\cos \left( \theta \right)  \left( {\frac {\partial }{\partial 
\theta}}H \left( r,x,\theta \right)  \right) {c}^{8}{k}^{8}{r}^{4}\nn\\
&+&64
\,\cos \left( \theta \right)  \left( {\frac {\partial }{\partial 
\theta}}H \left( r,x,\theta \right)  \right) {c}^{4}{k}^{4}{r}^{8}+
1024\,\sin \left( \theta \right)  \left( {\frac {\partial ^{2}}{
\partial {\theta}^{2}}}H \left( r,x,\theta \right)  \right) {c}^{8}{k}
^{8}{r}^{4}\nn\\
&-&64\,\sin \left( \theta \right)  \left( {\frac {\partial ^{
2}}{\partial {\theta}^{2}}}H \left( r,x,\theta \right)  \right) {c}^{4
}{k}^{4}{r}^{8}+1024\,\sin \left( \theta \right)  \left( {\frac {
\partial ^{2}}{\partial {\theta}^{2}}}H \left( r,x,\theta \right) 
 \right) {c}^{8}{k}^{4}{r}^{4}.\nn\\
 &&
\end{eqnarray}
Using the separation of variables $H(t,r,\theta)=\Theta(\theta)+R(r)X(x)$, we find ${\cal L}_2=0$ leads to
\be
\Theta(\theta)=\xi \cos\theta+\zeta,
\ee
where $\xi$ and $\zeta$ are two constants. Equation ${\cal L}_1=0$ leads to
\be
{\cal P}_1(r)\frac{1}{R(r)}\frac{d^2R(r)}{dr^2}+{\cal P}_2(r)\frac{1}{R(r)}\frac{dR(r)}{dr}+{\cal Q}_1\frac{1}{X(x)}\frac{d^2X(x)}{dx^2}+\frac{{\cal Q}_2(r,x,\theta)}{R(r)X(x)}=0,\label{irsepeq}
\ee
where
\be
{\cal P}_1(r)={\frac { \left( {r}^{14}-16\,{c}^{4} \left( {k}^{4}+1 \right) {r}^{10}
+256\,{c}^{8}{k}^{4}{r}^{6} \right) \sqrt {F \left( r \right) }}{16384({r}^{14}-16\,{c}^{4} \left( {k}^{4}+1 \right) {r}^{10}+256\,{c
}^{8}{k}^{4}{r}^{6})
}},
\ee
\begin{eqnarray}
{\cal P}_2(r)=
\frac {{r}^{5} \left(  \left( {\frac 
{{r}^{9}}{256}}-1/16\,{c}^{4} \left( {k}^{4}+1 \right) {r}^{5}+{c}^{8}
{k}^{4}r \right) {\frac {\rm d}{{\rm d}r}}F \left( r \right)
 +7\,F
 \left( r \right)  \left( {\frac {3\,{r}^{8}}{1792}}-{\frac {5\,{c}^{4
} \left( {k}^{4}+1 \right) {r}^{4}}{112}}+{c}^{8}{k}^{4} \right) 
 \right) }{64\,
\sqrt {F \left( r \right) } ({r}^{14}-16\,{c}^{4} \left( {k}^{4}+1 \right) {r}^{10}+256\,{c
}^{8}{k}^{4}{r}^{6})}, \nonumber\\
&&
\end{eqnarray}
\be
{\cal Q}_1={\frac {1}{16384 }},
\ee
and
\be
{\cal Q}_2(r,x,\theta)={\frac { \xi \left( 2048\,{}\,{c}^{12}{k}^{8}-128\,{}\,{k}
^{4}{r}^{4} \left( {k}^{4}+3 \right) {c}^{8}+24\,{r}^{8}{}\,
 \left( {k}^{4}+2/3 \right) {c}^{4}-{}\,{r}^{12} \right) \cos
 \left( \theta \right) }{2048\,\sqrt {F \left( r \right) }X \left( x
 \right) R \left( r \right)({r}^{14}-16\,{c}^{4} \left( {k}^{4}+1 \right) {r}^{10}+256\,{c
}^{8}{k}^{4}{r}^{6})}}.
\ee
Clearly the differential equation (\ref{irsepeq}) is not separable, due to the non-zero function ${\cal Q}_2(r,x,\theta)$. This term is zero only if $\xi=0$, which points that the metric function $H$ can't be a function of $\theta$. Of course, proceeding with $H(r,\theta)$ to solve the field equations, leads to the already known solutions, as given in (\ref{genfin}), along with $\Lambda=0$.  We also considered the  metric function depending on other spatial directions. However, we always find that we can consistently solve all the field equations only if $H$ is a function of $r$ and $x$, which in turn leads to $\Lambda=0$.  Moreover, we considered other choices for the ansatz (\ref{ds6}) with arbitrary exponents for the metric function. However the field equations imply that the exponents should be as $-2$ and $\frac{2}{D-3}$, as we considered in all ansatzes through the article.
}}

\end{document}